\documentclass[a4paper,fleqn]{cas-sc}

\usepackage[utf8]{inputenc}
\usepackage[numbers]{natbib}
\usepackage{textcomp}
\usepackage{longtable}
\usepackage{calc}
\usepackage{graphicx}
\graphicspath{{figures/}}
\usepackage{listings}

\newcommand{\FixXPT}{\textit{FixXPT}}      % this program
\newcommand{\pyxPT}{\textit{py-xPT}}       % the stand-alone post-processor
\newcommand{\code}[1]{\texttt{#1}}         % C++ identifiers, calls, file names
\newcommand{\kw}[1]{\texttt{#1}}           % LAMMPS input-script keywords/values

\lstdefinestyle{fixxpt}{
  basicstyle=\ttfamily\footnotesize,
  commentstyle=\itshape\color{black!60},
  keywordstyle=\bfseries,
  showstringspaces=false,
  columns=fullflexible,
  keepspaces=true,
  breaklines=true,
  breakatwhitespace=true,
  breakindent=2em,
  aboveskip=0.8\baselineskip,
  belowskip=0.6\baselineskip,
  xleftmargin=1em,
}
\lstdefinelanguage{LAMMPS}{
  morekeywords={fix,unfix,compute,thermo_style,thermo,run,units,atom_style,
                velocity,pair_style,pair_coeff,timestep,variable},
  morecomment=[l]{\#},
  sensitive=true,
}

\begin{document}

\let\WriteBookmarks\relax
\def\floatpagepagefraction{1}
\def\textpagefraction{.001}

\shorttitle{On-the-fly Quantum-Corrected Thermodynamics in LAMMPS}
\shortauthors{Ramji et~al.}

\title[mode=title]{On-the-fly Estimation of Quantum-Corrected Thermodynamics in LAMMPS}

\author[1]{Robert Ramji}[orcid=0000-0002-9850-7138]
\affiliation[1]{organization={ATLAS Materials Physics Laboratory, Aiiso Yufeng Li Family Department of Chemical and NanoEngineering, University of California San Diego},
            city={La Jolla},
            citysep={,},
            postcode={92093},
            state={CA},
            country={USA}}

\author[1]{Ilya Mazalov}[orcid=0009-0008-5999-0981]

\author[2]{Ricardo Buarque}[orcid=0009-0000-9334-7190]
\affiliation[2]{organization={Program in Materials Science and Engineering, University of California San Diego},
            city={La Jolla},
            citysep={,},
            postcode={92093},
            state={CA},
            country={USA}}
            
\author[1]{Alexandria Do}[orcid=0000-0002-9124-6502]

\author[1]{Xiaoxu Ruan}[orcid=0009-0009-9897-4105]

\author[1,2]{Tod A. Pascal}[orcid=0000-0003-2096-1143]
\cormark[1]
\ead{tpascal@ucsd.edu}
\cortext[1]{Corresponding author}

\begin{abstract}
We present a native implementation of the Two/Three-Phase Thermodynamics (xPT, $x \in \{2,3\}$)
method within the LAMMPS molecular dynamics framework, termed \FixXPT{}. xPT is an approach for obtaining absolute, quantum corrected thermodynamics (entropy, enthalpy, free energy, heat capacity, diffusivity) from the density of states function, obtained from the velocity
autocorrelation function of an equilibrium trajectory. \FixXPT{} addresses
non-trivial integration challenges, including atom reordering, compute
scheduling, FFT backends, unit systems, and partitioning of atomic
velocities into molecular modes. Two autocorrelators are provided: a
Wiener--Khinchin FFT correlator with one cumulative window-end
operation, and an in-line multi-tau correlator whose memory footprint
grows only logarithmically with the window length, removing the
principal scaling bottleneck for long analysis windows. The velocity
history of each window is distributed across the MPI ranks, so the memory
the fix adds to a node stays nearly constant as ranks are added and its
cost falls with the number of GPUs. Replaying identical velocity
frames through \FixXPT{} and the stand-alone \pyxPT{} post-processor
reproduces every thermodynamic observable of well-sampled groups to
within 0.1--0.5\% across the full analysis-mode
matrix; sparsely populated ionic species agree within their
statistical uncertainty, and the multi-tau VACF converges monotonically
to the FFT result as its buffer grows, from 0.8\% to 0.1\% drift across
a 16-fold buffer sweep.
The code is distributed as a drop-in LAMMPS package.
\end{abstract}

\begin{highlights}
\item Native LAMMPS fix computes on-the-fly 1PT/2PT/3PT entropy and free energy.
\item FFT and in-line multi-tau velocity autocorrelators, both validated.
\item Matches the py-xPT post-processor to 0.1--0.5\% across every mode and correlator.
\item Concurrent fixes share one velocity buffer, making a mode sweep near-free.
\item Distributed frame buffers scale the fix to 128 CPU ranks and 16 A100 GPUs.
\end{highlights}


\maketitle

\vspace{1em}
\noindent\rule{\linewidth}{0.5pt}
\begin{description}
\item[Program title:] FixXPT
\item[CPC Library link to program files:] (to be added by Technical Editor)
\item[Developer's repository link:] \url{https://github.com/atlas-nano/codes/tree/main/fix-xpt}
\item[Code Ocean capsule:] (to be added by Technical Editor)
\item[Licensing provisions:] MIT license (MIT)
\item[Programming language:] C++17
\item[Supplementary material:] None.
\item[Nature of problem:] Computing entropies and free energies from
  equilibrium molecular dynamics simulations is challenging because these
  quantities are not simple ensemble averages. Standard approaches such as
  free-energy perturbation and thermodynamic integration require multiple
  simulations and specialized sampling protocols. Post-processing velocity
  trajectories with stand-alone codes requires storing large trajectory
  files and precludes on-the-fly monitoring.
\item[Solution method:] We implement the Two/Three-Phase Thermodynamics (2PT/3PT)
  method~[1--3] as a native LAMMPS~[4] fix. The velocity autocorrelation
  function (VACF) is self-consistently partitioned into gas-like and solid-like
  components. Thermodynamic quantities follow from analytical integration
  of each component with appropriate weighting functions. The fix operates entirely
  in-memory, writing results to thermo output and summary files at the end
  of each analysis window without generating velocity trajectory files.
\item[Additional comments including Restrictions and Unusual features:]
  \emph{External routines.} The package builds within a LAMMPS source tree
  and requires an MPI implementation and an FFT backend. The
  optional GPU styles additionally require the LAMMPS KOKKOS
  package~[5]. \emph{Restrictions.} Buffer sharing between concurrent fixes requires the FFT correlator; the 3PT cage is
  skipped under LJ reduced units. The fix requires atom IDs and per-type
  masses. GPU results are not bit-equivalent to the CPU path: the device path
  is double precision throughout and differs only in summation order.
  \emph{Features.} By default the frame history is distributed across MPI
  ranks, so the per-rank memory falls inversely with the rank count. Concurrent fixes that accumulate the same
  trajectory share one velocity buffer, making subsequent sweeps nearly memory-free.
  Supports all eight LAMMPS unit systems.
  Molecular decomposition into translational, rotational, and vibrational
  contributions for polyatomic molecules. A selectable 1PT/2PT/3PT mode
  partition: the 1PT all-solid harmonic baseline; the 2PT gas/solid
  partition with four entropy refinements; and the 3PT parameter-free anharmonic cage extension.
  Compatible with any LAMMPS FFT backend.
\item[References:]
  \begin{enumerate}
  \def\labelenumi{[\arabic{enumi}]}
  \item S.-T.~Lin, M.~Blanco, W.~A.~Goddard~III, The two-phase model for
    calculating thermodynamic properties of liquids from molecular dynamics:
    Validation for the phase diagram of Lennard-Jones fluids, J. Chem. Phys.
    119 (2003) 11792--11805. \url{https://doi.org/10.1063/1.1624057}
  \item S.-T.~Lin, P.~K.~Maiti, W.~A.~Goddard~III, Two-phase thermodynamic
    model for efficient and accurate absolute entropy of water from molecular
    dynamics simulations, J. Phys. Chem. B 114 (2010) 8191--8198.
    \url{https://doi.org/10.1021/jp103120q}
  \item R.~Buarque, M.~Gascon, T.~A.~Pascal, An anharmonic liquid-entropy
    functional from the Mori--Zwanzig memory kernel, arXiv:2607.29026
    [physics.chem-ph] (2026). \url{https://doi.org/10.48550/arXiv.2607.29026}
  \item A.~P.~Thompson, H.~M.~Aktulga, R.~Berger, et al., LAMMPS -- a flexible
    simulation tool for particle-based materials modeling at the atomic, meso,
    and continuum scales, Comput. Phys. Commun. 271 (2022) 108171.
    \url{https://doi.org/10.1016/j.cpc.2021.108171}
  \item C.~R.~Trott, D.~Lebrun-Grandi\'e, D.~Arndt, et al., Kokkos 3:
    Programming model extensions for the exascale era, IEEE Trans. Parallel
    Distrib. Syst. 33 (2022) 805--817.
    \url{https://doi.org/10.1109/TPDS.2021.3097283}
  \end{enumerate}
\end{description}
\noindent\rule{\linewidth}{0.5pt}
\vspace{1em}

\section{Introduction}\label{introduction}

Entropy and free energy underpin the thermodynamic stability of condensed phases, solvation, phase transitions and equilibria, and biomolecular binding, yet they are difficult to compute from standard molecular dynamics (MD) simulations. Unlike observables such as energy, temperature, or pressure, which are obtained by straightforward ensemble averaging, free energies and entropies require integration over configuration space. Established and formally exact methods, such as free-energy perturbation~\cite{Zwanzig1954} and thermodynamic integration~\cite{Kirkwood1935}, achieve this through carefully designed thermodynamic paths connecting a reference state to the system of interest. This approach, however, demands multiple simulation runs and carries the statistical overhead of alchemical or collective-variable transformations.

The Two-Phase Thermodynamics (2PT) method of Lin, Blanco, and Goddard
~\cite{Lin2003} offers an alternative approach: a single, short
equilibrium MD trajectory can be sufficient to compute absolute
thermodynamic quantities of most condensed phase systems. The method
maps the density of states (DoS), obtained from the cosine transform of
the mass-weighted velocity autocorrelation function (VACF), onto an
analytically integrable model. The DoS is partitioned into 1) a gas-like
component, described by a hard-sphere Lorentzian parameterized by the
zero-frequency DoS value and 2) a solid-like component, whose
thermodynamics are obtained from the Debye vibrating solid model with quantum
harmonic oscillator weights. The key self-consistency parameter, the
fluidicity $f$, the fraction of degrees of freedom treated as gas-like,
is determined by equating the system's self-diffusivity to that of a
hard-sphere fluid with the same number of particles at the same temperature and density. The method was
subsequently extended to improve the
Sackur--Tetrode entropy term, applied to molecular systems,~\cite{Lin2010} where
the atomic VACF was decomposed into translational, rotational, and
vibrational contributions. We extended the approach to
describing the thermodynamics of liquid mixtures ~\cite{pascal2012hydrophobic}, and recently to include a cage entropy term built from the Mori-Zwanzig memory kernel ~\cite{Buarque2026}. 

Since its introduction, 2PT has been applied across a broad range of
condensed-matter problems, including thermodynamic properties of ionic liquids
~\cite{Endo2022}, phase diagrams of
metallic systems ~\cite{Desjarlais2013}, and entropy of water under confinement ~\cite{Pascal2011PNAS}.
In all of these applications, the standard workflow uses stand-alone
post-processors, written either in Python~\cite{Ramji2026pyxpt} or C++~\cite{Ramji20232pt}, applied to
velocity trajectories saved to disk. This approach has two practical
drawbacks: (1) velocity trajectory files for large systems and long
windows can be gigabytes to tens or hundreds of gigabytes, posing
storage and I/O bottlenecks; and (2) the approach is not available
during the simulation, preventing on-the-fly monitoring or adaptive
steering.

Here we describe \FixXPT{}, which addresses both drawbacks by integrating
xPT directly into LAMMPS ~\cite{Thompson2022} as a native fix. The name reflects the
selectable phase partitioning exposed through a \kw{mode} keyword: a one-phase all-solid baseline (1PT), the
standard two-phase gas/solid partition (2PT), and a three-phase cage
extension (3PT). Velocities are
accumulated in memory, the VACF and DoS are computed using the LAMMPS
FFT infrastructure, and thermodynamic results are written to output
files and the LAMMPS thermo vector at the end of each analysis
window. No trajectory files are written or read. Our implementation
supports all eight LAMMPS unit systems (\kw{real}, \kw{metal}, \kw{lj}, \kw{si},
\kw{cgs}, \kw{micro}, \kw{nano}, \kw{electron}) and accommodates dynamic groups,
enabling thermodynamic analysis of sub-regions (e.g., solvation shells,
binding sites, and interfacial layers) within a single continuous
simulation. We also provide a full implementation of the Desjarlais
memory-function model ~\cite{Desjarlais2013}, which replaces the Lorentzian gas DoS with
a physically motivated Gaussian-kernel that better captures the spectral
shape of structured liquids. To extend xPT to long analysis
windows, where the per-rank velocity buffer of an FFT-based correlator
becomes the dominant memory cost, we additionally provide an in-line
multi-tau correlator~\cite{ramirez2010efficient} that buffers only a small number of
telescoping samples, growing logarithmically with the window length, and
produces a VACF whose drift relative to the
full FFT path is below 1\% for sensible parameter choices.

The remainder of this paper is organized as follows: Section~\ref{theory}
summarizes the xPT theoretical framework. Section~\ref{molecular-2pt-decomposition} covers the
molecular decomposition mode. Section~\ref{code-architecture} describes the
overall code architecture and the LAMMPS fix lifecycle. Section~\ref{key-implementation-challenges}
details engineering challenges and their solutions. Section~\ref{verification} gives validation results.
Section~\ref{performance} discusses performance and memory usage. Section~\ref{conclusions} concludes. A
full description of the implementation of the Desjarlais memory function
model is provided in Appendix~\ref{app:desjarlais}.

\section{Theory}\label{theory}

\subsection{Density of States from the VACF and diffusivity}\label{density-of-states-from-the-vacf-and-diffusivity}

We form the mass-weighted velocity autocorrelation function

\begin{equation}
C(t) = \sum_i m_i \langle \mathbf{v}_i(0)\cdot\mathbf{v}_i(t)\rangle
\end{equation}
where \(m_{i}\) is the mass of atom \(i\) and the angle brackets denote a
time average. Equipartition fixes its zero-lag value at $C(0) = 3Nk_BT$, and
the normalized form $c(t) = C(t)/C(0)$ is the quantity the memory-kernel
analysis operates on. The vibrational DoS is the one-sided cosine transform
of $C(t)$:
\begin{equation}\label{eq:dos}
g(\nu) = \frac{4}{k_BT}\int_0^{\infty} C(t)\cos(2\pi\nu t)\,dt,
\qquad \int_0^{\infty} g(\nu)\,d\nu = 3N
\end{equation}
normalized to the $3N$ translational degrees of freedom.

In practice, the DoS is computed from a finite window of \(N_{s}\) frames
sampled every \(N_{\text{every}}\) MD steps, giving a frequency resolution
of $\Delta\nu = \frac{1}{2N_s\Delta t_{vac}}$ where $\Delta t_{vac} = N_{every}\Delta t_{MD}$. The zero-frequency
value $g(0) \equiv S_0$ is related to the self-diffusivity $D$
through the relation:
\begin{equation}
S_0 = \frac{4}{k_BT}\int_0^{\infty} C(t)\,dt = \frac{12mND}{k_BT}
\end{equation}

\subsection{2PT Partitioning and Fluidicity}\label{pt-partitioning-and-fluidicity}

The total DoS is partitioned into additive gas and solid components,
$g(\nu) = g_{gas}(\nu) + g_{solid}(\nu)$, where the dimensionless
fluidicity $f \in [0,1]$ is the fraction of the $3N$ translational
degrees of freedom that are treated as gas-like, so that
$N^{dof}_{gas} = 3fN$ for a group of $N$ atoms.  
In the standard
implementation, the gas component takes the form
\begin{equation}
g_{gas}(\nu) = \frac{S_0}{1+\left(\frac{\pi S_0 \nu}{6fN}\right)^2}
\end{equation}
which is a Lorentzian pinned to the zero-frequency intensity,
$g_{gas}(0) = S_0$, and normalized so that
$\int_0^{\infty} g_{gas}(\nu)\,d\nu = N^{dof}_{gas} = 3fN$. The solid component $g_{solid} = g - g_{gas}$ contains the
remaining modes.

In practice, $f$ is determined by relating the Green-Kubo self-diffusivity of the system to that of a hard-sphere fluid with the same number of particles, at the same temperature and density, through
Chapman--Enskog theory~\cite{Chapman1970} with the hard-sphere packing fraction $y$
supplied by the Carnahan--Starling equation of state~\cite{Carnahan1969}. We define the normalized,
dimensionless diffusivity parameter $\Delta$:
\begin{equation}
\Delta = \frac{S_0}{N}\sqrt{\frac{\pi k_B T}{m}}\ \frac{2}{9}\left(\frac{N}{V}\right)^{1/3}\left(\frac{6}{\pi}\right)^{2/3}
\end{equation}
which sets $y$ and, by equivalence, $f$:
\begin{equation}
y = \frac{f^{5/2}}{\Delta^{3/2}}
\qquad\Longleftrightarrow\qquad
f = \Delta^{3/5}\,y^{2/5}
\end{equation}
Thus the diffusivity requirement produces the
Lin--Blanco--Goddard fluidicity equation, a polynomial in \(f^{1/2}\)
(equivalently a higher-order polynomial in $f$) parameterized by \(\Delta\):
\begin{equation}
2\Delta^{-4.5}f^{7.5} - 6\Delta^{-3}f^{5} - \Delta^{-1.5}f^{3.5} + 6\Delta^{-1.5}f^{2.5} + 2f - 2 = 0
\end{equation}
The physical root is found by Newton--Raphson iteration using the
starting estimate $f_0 = 0.7293\Delta^{0.5727}$ ~\cite{Lin2003}.

\subsection{Thermodynamic Weighting Functions}\label{thermodynamic-weighting-functions}

Once \(g(\nu)\) is partitioned, thermodynamic quantities are obtained by
integrating each component with quantum weighting functions. For the
solid component, the Debye vibrating crystal model with quantum harmonic weights produces the
entropy $S$:
\begin{equation}
\frac{S_{solid}}{k_B} = \int_0^{\infty} \left[\frac{u}{e^u-1}-\ln(1-e^{-u})\right]g_{solid}(\nu)d\nu,\quad u=\frac{h\nu}{k_BT}
\end{equation}
with analogous integrals for the Helmholtz free energy $A_{solid}$, and
the internal energy $E_{solid}$. For the gas component, the hard-sphere
(Lorentzian) density of states decays as $\nu^{-2}$ with frequency $\nu$, and we
compute the entropy by integrating the gas DoS against a
frequency-independent per-mode weight, built from the Sackur--Tetrode
entropy of an ideal gas $S^{IG}$ and the Carnahan--Starling excess:
\begin{equation}
\frac{S_{gas}}{k_B} = \frac{1}{3}\left[\frac{S^{IG}}{k_B}+\ln Z_{\mathrm{HS}}(y)+\frac{y(3y-4)}{(1-y)^2}\right]\int_0^{\infty} g_{gas}(\nu)\,d\nu
\end{equation}
Of note: (a) the $1/3$ factor arises because the weight is per translational degree of freedom while the bracket is a
per-particle entropy, and $\int g_{gas}\,d\nu = 3fN$ counts modes; and (b) the
$\ln Z_{\mathrm{HS}}(y)$ term is an ansatz of the original derivation~\cite{Lin2003}, introduced to obtain results more consistent with free-energy perturbation, and it deviates from the canonical definition.

To obtain the energy we use a constant weight of 0.5, while the
Helmholtz free energy assumes a weight that is the difference between
the energy and entropy weights. 
For pure, single-component systems, the chemical potential is reported as $\mu = (A + PV)/N$, where the translational gas-phase $PV$ term is $R T\, Z_{\mathrm{HS}}(y)/3$ per gas-like degree of freedom, with $Z_{\mathrm{HS}}(y)$ the hard-sphere compressibility factor at the packing fraction $y$. Note that for multi-component mixtures or dynamic sub-groups, this computed quantity formally represents the species' absolute molar Gibbs free energy ($g_i = G_i/N_i$) rather than the true partial molar chemical potential $\mu_i$, as the 2PT method does not explicitly capture the combinatorial entropy of mixing. For molecular systems, the separate rotational gas is treated as an ideal rotor, for which $PV$ per particle is $k_B T$ irrespective of packing, contributing $R T/n_{\mathrm{rot}}$ per rotational gas-like degree of freedom. The solid contribution to $\mu$ (or $g_i$) equals its contribution to $A$.
The Helmholtz free energy itself is always
\(A = E - TS\); only \(\mu\) carries the
compressibility contribution. Purely classical energies (i.e., replacing quantum weights with
their high-$T$ limits) are computed in parallel and used for energy and
heat capacity corrections (see below). All integrals are evaluated by the
trapezoid rule on the discrete DoS array.

\subsection{Energy Correction}\label{energy-correction}

The DoS-derived classical energy \(E_{CHO}\) (i.e., the energy of the
system if all the solid modes were a classical harmonic oscillator)
differs from the running MD average $E_{MD} = \left\{KE+PE\right\}$ accumulated
over the window. Quantum
corrections are then obtained from evaluating the excess
energy using quantum harmonic oscillator weights $E_{QHO}$:
\begin{equation}
E_q = E_{MD} + \left(E_{QHO} - E_{CHO}\right);\ A_q = E_q - T S_q
\end{equation}
Similarly, the quantum heat capacity is obtained using the classical
term (from the total energy-fluctuation estimator
$\sigma_E^{2}$), and the excess term:
\begin{equation}
C_V = \frac{\sigma_E^{2}}{k_B T^{2}} + \left(C_{V,QHO} - C_{V,CHO}\right)
\end{equation}

\subsection{The Desjarlais Memory-Function Gas DoS}\label{the-desjarlais-memory-function-gas-dos}

The Lorentzian gas component is the solution to the generalized Langevin
equation for the VACF with a Markovian (delta-function) memory kernel.
Desjarlais ~\cite{Desjarlais2013} showed that replacing the delta-function with a
Gaussian memory kernel of bandwidth \(B_{g}\) yields a substantially
improved description of the gas DoS in structured liquids such as liquid
metals, where the VACF exhibits sub-diffusive behavior not captured by
the simple Lorentzian. The memory-function gas DoS is (see
Appendix~\ref{app:desjarlais} for the derivation):
\begin{equation}
g_{gas}^{HS}(\nu,B_g) = \dfrac{A_g\sqrt{\pi/(4B_g)}\,e^{-y^2}f}{(A_g)^2\dfrac{\pi}{4B_g}e^{-2y^2}+\dfrac{(A_g)^2D^2(y)}{B_g}-4A_g y D(y)+\tilde{\nu}^{2}}
\end{equation}
where $\tilde{\nu}=\pi\nu/6N,\ y=\tilde{\nu}/\sqrt{4B_g},\ D(y)=e^{-y^2}\int_0^{y}e^{t^2}\,dt$ is the Dawson integral, and
$A_g = \left(f/S_0\right)\sqrt{4B_g/\pi}$ is fixed by the constraint
$g_{gas}^{HS}(0) = S_0$. Note that as $B_g$ $\rightarrow$ $\infty$ the Gaussian kernel
collapses to a delta function and the expression reduces to the standard
Lorentzian (the Markovian limit; see Appendix~\ref{app:desjarlais}). In practical
implementations, the shape parameter $B_g$ is determined
self-consistently by matching $g_{gas}^{HS}(\nu)$ to the actual
computed DoS at the last frequency bin above the noise floor. In our implementation, the value
$B_g$ = 0 is reserved as a sentinel that signals a failed fit, in which
case the code falls back to the Lorentzian gas DoS.

\subsection{The Three-Phase (3PT) Cage}\label{the-three-phase-3pt-cage}

The 2PT gas/solid partition treats every non-diffusive mode as a
harmonic solid, which neglects the anharmonic entropy of the transient
``cage'' a particle explores before it either diffuses away or relaxes
harmonically. The three-phase (3PT) extension~\cite{Buarque2026} adds this cage as an
explicit third component built from the non-Markovian memory of the
velocity autocorrelation. Writing the friction (memory) kernel $K(t)$
of the generalized Langevin equation as a Volterra integral of the VACF,
the cage density of states is the excess of the true mobility
\(F_{K}(\nu) = \mathrm{Re}[1/(i\omega + \tilde{K}(i\omega))], \quad \omega = 2\pi \nu\) over its Markovian
(Lorentzian) counterpart \(F_{M}(\nu) = \gamma/(\gamma^{2}+\omega^{2})\),
with \(\gamma = \tilde{K}(0)\). The corresponding entropy correction is
integrated with the gas--solid weight difference under a per-degree-of-
freedom prefactor and a fluidicity gate, and applied to the
translational and (for molecules) rotational channels. The cage term requires no additional
parameters and vanishes exactly in the crystalline and
ideal-Markovian limits. Its construction, sum rules, and 
validation against free-energy-perturbation entropies are developed in
reference ~\cite{Buarque2026}.

\section{Molecular xPT Decomposition}\label{molecular-2pt-decomposition}

\subsection{Velocity Projection}\label{velocity-projection}

For polyatomic molecules, the atomic velocity is decomposed as:
\begin{equation}
\mathbf{v}_i = \mathbf{v}_{COM} + \boldsymbol{\omega} \times \mathbf{r}_i' + \mathbf{v}_{vib,i}
\end{equation}
where \textbf{v}\textsubscript{COM} is the molecular center-of-mass velocity, $\omega$ is the
angular velocity, and r'$_i$ = \textbf{r}\textsubscript{i} - \textbf{r}\textsubscript{COM} uses unwrapped
coordinates to avoid periodic-image artifacts. For each frame and each
molecule, the fix computes:
\begin{enumerate}
\def\labelenumi{\arabic{enumi}.}
\item
  \textbf{COM velocity} $\mathbf{v}_{COM} = \left(\sum_i m_i \mathbf{v}_i\right)/M_{mol}$ (one MPI reduce).
\item
  \textbf{Angular momentum} $\mathbf{L} = \sum_i m_i \mathbf{r}_i' \times \left(\mathbf{v}_i - \mathbf{v}_{COM}\right)$ and inertia tensor \textbf{I}
  (second MPI reduce).
\end{enumerate}
The angular velocity $\boldsymbol{\omega} = \mathbf{I}^{-1} \mathbf{L}$ is computed in the principal
frame of \textbf{I}. For linear molecules, the principal axis with the
smallest moment is excluded by setting $\omega$ = 0 along that axis, giving 2
rotational DOFs. The same guard is applied in the hard-sphere rotational
entropy: the vanishing axial moment would otherwise produce a divergent
rotational temperature, so the finite perpendicular moment is used
throughout for linear species.

\subsection{Principal-Frame Diagonalization}\label{principal-frame-diagonalization}

Diagonalization of the $3 \times 3$ symmetric inertia tensor uses a
self-contained Jacobi iteration (\code{mat3\_sym\_eigen}), requiring no
external linear-algebra library. In at most 50 sweeps over the three
off-diagonal pairs $\{(0,1),\ (0,2),\ (1,2)\}$, the algorithm converges to
machine precision for typical inertia tensors. Eigenvalues are sorted
ascending; eigenvectors accumulate in the columns of the rotation matrix
\textbf{V}.
For the rotational VACF, the scaled angular velocity
$\boldsymbol{\omega} = \left(L_1/\sqrt{I_1},\ L_2/\sqrt{I_2},\ L_3/\sqrt{I_3}\right)$ is stored in the body frame and then rotated back
to the lab frame. This square-root inertia weighting gives the
rotational VACF the same mass-weighted structure as the translational
VACF (units of energy per molecule), so the same spectral integration
routines serve both.

\subsection{Rotational Hard-Sphere Entropy}\label{rotational-hard-sphere-entropy}

The gas-phase rotational entropy is computed in a similar fashion to the
gas-phase translation entropy: integrating over the rotational DoS using
weights obtained from the quantum rigid-rotor partition function,
weighted by the rotational fluidicity $f_{rot}$. For a nonlinear molecule
with principal moments I\textsubscript{1}$\leq$I\textsubscript{2}$\leq$I\textsubscript{3} and symmetry number $\sigma_{rot}$, the
weights are:
\begin{equation}
\frac{S_{gas}^{rot}}{k_B} = \frac{1}{n_{rot}}\left\{\frac{3}{2} + \ln\left[\frac{\sqrt{\pi}}{\sigma_{rot}}\sqrt{\frac{\left(8\pi^{2} k_B T\right)^{3} I_1 I_2 I_3}{h^{6}}}\right]\right\} \int_0^{\infty} g_{gas}^{rot}(\nu)\,d\nu
\end{equation}
Here $n_{rot}$ is the number of active rotational degrees of freedom,
three for a nonlinear molecule and two for a linear one, so that the
braced term is again a per-mode weight. For linear molecules the
principal moment about the molecular axis vanishes (the code drops the
smallest-moment axis, per Section~\ref{velocity-projection}), and the
formula reduces to the diatomic rigid-rotor expression.

\section{Code Architecture}\label{code-architecture}

\subsection{The LAMMPS Fix Lifecycle}\label{the-lammps-fix-lifecycle}

\FixXPT{} derives from LAMMPS's \code{Fix} base class and integrates into the
simulation loop through four lifecycle callbacks:
\begin{longtable}[]{@{}
  >{\raggedright\arraybackslash}p{(\columnwidth - 2\tabcolsep) * \real{0.2278}}
  >{\raggedright\arraybackslash}p{(\columnwidth - 2\tabcolsep) * \real{0.7468}}@{}}
\toprule\noalign{}
\begin{minipage}[b]{\linewidth}\raggedright
Callback
\end{minipage} & \begin{minipage}[b]{\linewidth}\raggedright
Purpose
\end{minipage} \\
\midrule\noalign{}
\endhead
\bottomrule\noalign{}
\endlastfoot
\code{init()} & Resolve computes, build molecular topology, open output
file streams\\
\code{setup(vflag)} & Schedule the first energy/pressure tally via \code{addstep()} \\
\code{end\_of\_step()} & Accumulate one velocity frame; trigger analysis when
window is full \\
\code{post\_run()} & Flush a partial window if $\geq$ 25 \% of frames were collected (multi-tau correlator only) \\
\end{longtable}
%% cas-sc steps the table counter even for an un-captioned longtable;
%% this one is presented inline, so give the number back to Table 1.
\addtocounter{table}{-1}
The fix produces a thermo vector (\code{compute\_vector}) of 10 scalars in
monatomic mode, 14 in molecular mode, and 18 with the \kw{show\_split}
option, making all xPT results accessible to \kw{thermo\_style custom f\_ID[n]}. Each length comprises a core block of 8, 12, or 16 scalars
plus the quantum and classical chemical potentials \(\mu_{q}\) and \(\mu_{c}\).

\textbf{Fix syntax:}
\begin{lstlisting}[language=LAMMPS]
fix ID group-ID xpt Nevery Nframes prefix [keyword value ...]
\end{lstlisting}
Keywords:
\begin{itemize}
\item
  \kw{classical}: output classical alongside quantum thermodynamics
\item
  \kw{normalize}: per-atom/molecule output
\item
  \kw{epsilon E}, \kw{sigma S}, \kw{mass M}: LJ physical parameters for $\hbar$*
  scaling
\item
  \kw{volume V}: override simulation cell volume
\item
  \kw{molecule}: enable the molecular decomposition, 
  partitioning the entropy into translational, rotational, and
  vibrational channels; requires \kw{atom\_style molecular} or \kw{full}.
  Absent this keyword the fix runs in monatomic mode on the full DoS.
\item
  \kw{symmetry sym}: Schoenflies point-group symbol (default \kw{C1}) from
  which the molecular rotational symmetry number \(\sigma_{\text{rot}}\) is
  auto-derived (e.g.~\kw{C2v} $\rightarrow$ 2 for water, \kw{D6h} $\rightarrow$ 12 for benzene).
\item
  \kw{mode 1PT\textbar2PT\textbar3PT} (applies to both monatomic and molecular
  systems): selects the phase partition.
  \begin{itemize}
  \item
    \kw{1PT}: all-solid baseline, in which every degree of freedom is treated as solid-like, the fluidicity is set to zero and no gas DoS is computed.
  \item
    \kw{2PT} (default): the gas/solid hard-sphere partition of Lin,
    Blanco, and Goddard.
  \item
    \kw{3PT}: the three-phase cage extension, which augments the
    gas/solid partition with a parameter-free anharmonic cage-memory
    component (Section~\ref{the-three-phase-3pt-cage}).
  \end{itemize}
\item
  \kw{refinement rigorous\textbar lin2003\textbar desjarlais\textbar r2pt}: selects the 2PT
  sub-method (\kw{mode} = 2PT only):
  \begin{itemize}
  \item
    \kw{rigorous} (default): the thermodynamically exact hard-sphere
    reference entropy.
  \item
    \kw{lin2003}: the Lin--Blanco--Goddard 2003 convention, which adds
    the \(+\ln Z\) hard-sphere compressibility term.
  \item
    \kw{desjarlais}: replaces the Lorentzian gas DoS with the Desjarlais
    memory-function form (Section~\ref{the-desjarlais-memory-function-gas-dos}, Appendix); adopts the
    \kw{lin2003} \(+\ln Z\) entropy convention.
  \item
    \kw{r2pt}: the revised-2PT correction of Sun \emph{et al.} ~\cite{Sun2017}.
  \end{itemize}
\item
  \kw{correlator fft\textbar multitau}: select the velocity autocorrelator.
  \kw{fft} (default) accumulates the full $N_s$-frame velocity buffer
  and computes the VACF via one Wiener--Khinchin FFT at window end.
  \kw{multitau} runs an in-line telescoping (multi-tau) correlator~\cite{ramirez2010efficient} whose
  memory footprint grows only logarithmically with $N_s$ (Sections~\ref{on-line-multi-tau-correlator} and~\ref{memory-usage}). Its
  shape is controlled by \kw{mt\_M N} (base lag bins per level, default 32),
  \kw{mt\_P N} (inter-level overlap bins, default 16), and \kw{mt\_S N}
  (inter-level decimation factor, default 2); the number of levels is
  chosen automatically from $N_s$.
\item
  \kw{buffer\_layout distributed\textbar replicated}: where the frame
  history lives. \kw{distributed} (default) gives every group atom one
  home rank per window, so a rank stores about $1/P$ of the history on
  $P$ ranks (Section~\ref{tag-based-parallel-atom-migration}); \kw{replicated} keeps the complete history
  on every rank. On a single rank the two hold the same history, and the fix
  selects \kw{replicated} for atomic groups, where it is faster
  (Section~\ref{gpu-scaling}); an explicit setting always wins.
\end{itemize}

\subsection{Data Layout}\label{data-layout}

The central data structure is a three-dimensional velocity buffer,
\code{vel\_buf[frame][slot][dim]} with $\mathrm{dim} \in \{x,y,z\}$, whose slots are the
atoms a rank holds. In the default distributed layout a rank holds its
home atoms, about $N/P$ of the group (Section~\ref{tag-based-parallel-atom-migration}), so the monatomic
buffer holds $N_s \times \lceil N/P \rceil \times 3$ doubles per rank; in the
replicated layout every rank holds the whole group at
$\mathrm{slot} = \mathrm{atom\_tag} - 1$, and the buffer holds $N_s \times N_{\mathrm{max\_tag}} \times 3$
doubles. In molecular mode, four additional frame buffers carry the
per-molecule center-of-mass velocity (\code{com\_vel\_buf}), scaled angular
velocity (\code{omega\_buf}) and lab-frame angular velocity (\code{angmom\_buf}), and the
per-atom vibrational residual (\code{vib\_vel\_buf}).

\subsection{Shared Velocity Buffers}\label{shared-velocity-buffers}
The memory buffer is the dominant memory cost of the fix. A common use case for \FixXPT{} is to analyze the same trajectory using multiple modes to, for example, evaluate the anharmonic contributions to the thermodynamics from 1PT vs 2PT. Declaring one fix per setting is the
natural way to do this, and it would ordinarily require one
complete velocity history per setting and stage every frame into each of
them.
Operating on the insight that the trajectory a fix accumulates does not depend on what it
subsequently does with it, \FixXPT{} detects when a buffer is reusable
and instead shares it. At each \code{init()} it scans the LAMMPS fix list for
an earlier \kw{xpt} fix whose configuration is equivalent, requiring the
same group, the same $N_s$ and $N_{every}$, the same correlator, the same
buffer precision and layout, and the same molecular setting. \kw{mode} and \kw{refinement} are deliberately
absent from that key: both act on the density of states, downstream of
the trajectory. 

A fix that finds a partner becomes a \emph{consumer}. It
re-points its buffer pointers at the owner, is re-aliased whenever the
owner grows the buffer, and skips frame accumulation altogether.
A consumer therefore adds no memory, no per-frame velocity staging and no
per-frame communication. It pays only for its own window-end
analysis, which is the part that differs between settings. On the
six-setting argon sweep distributed with the package (500 atoms,
$N_s = 4096$, \code{np} = 1), six concurrent fixes peak at 104.5 MB
against 102.9 MB for
a single fix: five further analyses cost 1.6 MB. (The single-fix
baseline is consistent with the argon row of Table~\ref{tab:15}: its
42.8 MB at $N_{s} = 512$ plus the $\approx$41 MB of buffer growth to
$N_{s} = 4096$ and the associated FFT workspace.) Placing each fix on its
own group of the same atoms defeats the sharing and raises the peak to
424.7 MB, an increment of 319.8 MB --- $\approx$64 MB per redundant
velocity history, the $4096 \times 500 \times 3 \times 8$-byte buffer
plus its FFT workspace. The six-way sweep of
Section~\ref{complete-mode-correlator-sweep} therefore costs one velocity
history rather than six. Detection runs once per \kw{run} command, so
fixes added or deleted between runs are handled without user
intervention.
This feature is enabled only for the FFT correlator,
because the multi-tau ring buffers are per-fix and a multi-tau consumer
would reach the end of its window with no autocorrelation of its own;
multi-tau fixes always own their buffers.

\subsection{Analysis Pipeline}\label{analysis-pipeline}

At the end of each full window \code{run\_analysis()} executes the following
steps:

\begin{enumerate}
\def\labelenumi{\arabic{enumi}.}
\item
  \textbf{FFT setup.} Instantiate a 1-D \code{FFT3d} object of length $N_{FFT} = 2N_s$
\item
  \textbf{Local power spectrum.} For each atom the rank holds (its home
  atoms in the distributed layout) and each Cartesian direction, copy the $N_s$-point velocity signal into a zero-padded
  buffer of length $2N_s$, forward-FFT, and accumulate the
  mass-weighted $\left|S_{\nu}\right|^{2}$ into \code{pwr\_local}.
\item
  \textbf{MPI reduction.} \code{MPI\_Allreduce(SUM)} across all ranks to
  obtain the global power spectrum
\item
  \textbf{IFFT $\rightarrow$ VACF.} Backward FFT of the power spectrum; divide by
  $N_{FFT}\,N_s$ to obtain the normalized VACF. When
  \kw{correlator multitau} is selected (Section~\ref{on-line-multi-tau-correlator}), steps 1--4 are
  replaced by an in-line multi-tau accumulator ~\cite{ramirez2010efficient} that updates the VACF at every accepted
  frame and emits the normalized VACF on a logarithmically spaced
  lag grid at window end, which is then resampled onto the uniform
  time grid expected by step 5; no large $2N_s$ velocity buffer is
  required.
\item
  \textbf{Cosine DoS.} Zero-symmetric cosine transform of the normalized
  VACF (period $2N_s$) to obtain the one-sided DoS on $N_s + 1$ bins
\item
  \textbf{Fluidicity.} Compute $\Delta$, solve for $f$ via Newton--Raphson in
  \code{search2pt}.
\item
  \textbf{Desjarlais $B_g$} (2PT-Des only). Determine $B_g$ by
  bisection (Appendix~\ref{app:desjarlais}) in \code{refine\_Bg\_des}
\item
  \textbf{Thermodynamic integration.} Trapezoid quadrature over
  $g_{solid}$ and $g_{gas}$ with quantum and classical weights
\item
  \textbf{Energy correction.} Apply excess energy $E_0$ and the
  fluctuation $C_v$ corrections
\item
  \textbf{Output.} Write a summary row to \code{prefix.thermo}, a per-window
  DoS block to \code{prefix.pwr}, and a VACF block to \code{prefix.vac} (the
  multi-tau correlator additionally emits \code{prefix.vac.multitau});
  update the LAMMPS thermo vector.
\end{enumerate}
In molecular mode, steps 2--5 are repeated for translational,
rotational, and vibrational velocity components.

\subsection{In-line Multi-tau Correlator}\label{on-line-multi-tau-correlator}

The FFT correlator of Section~\ref{analysis-pipeline} must retain the full $N_s$-frame
velocity buffer until window end, so its per-rank memory grows linearly
with the window length (Section~\ref{memory-usage}). For systems whose slowest relaxation
exceeds the affordable buffer (supercooled and glassy liquids, polymer melts, associated fluids), this buffer, not the force evaluation, becomes
the binding constraint. \FixXPT{} therefore provides an alternative in-line
multi-tau correlator~\cite{Schaetzel1987,Magatti2001,ramirez2010efficient}
(selected with \kw{correlator multitau}) that estimates
the VACF directly during the run with a memory footprint that grows only
logarithmically with $N_s$.

The multi-tau (or ``telescoping block'') correlator organizes the lag axis into a
hierarchy of $L$ levels of increasing time resolution. Level 0 accumulates
the products \(v(t)\, v(t-\tau)\) at the native sample interval
\(\Delta t_{\mathrm{vac}}\) for a block of \(M + P\) linearly spaced lags
(\kw{mt\_M} base bins plus \kw{mt\_P} overlap bins). The level-0 signal is then
block-averaged in groups of $S$ consecutive frames (\kw{mt\_S}, the decimation
factor, default 2) and fed to level 1, which accumulates the same
\(M + P\) lags at the coarser spacing \(S\,\Delta t_{\mathrm{vac}}\); level
\(\ell\) resolves lags at spacing \(S^{\ell}\,\Delta t_{\mathrm{vac}}\). The
$P$ overlap bins at the top of each level coincide with the first lags of
the next, so the per-level blocks splice into a single, smooth,
logarithmically spaced VACF spanning lags from \(\Delta t_{\mathrm{vac}}\)
to \(\sim N_{s}\,\Delta t_{\mathrm{vac}}\). The number of levels is set by
\kw{mt\_L}, or chosen automatically as
\(L = \lceil \log_{S}\!\left(N_{s}/(M{+}P)\right) \rceil\).

Each accepted frame triggers one push: level 0 is updated unconditionally,
and level \(\ell\) is updated once every \(S^{\ell}\) frames as its
decimation buffer fills. The per-frame cost is therefore
$\mathcal{O}\!\left((M+P)\,L\right)$ per channel, a small, fixed amount of work
that replaces the once-per-window FFT of the batch path with a steady
per-step push. The state retained between frames comprises the
\((M+P)\times L\) array of running lag sums together with, at each level,
a ring buffer holding the last \(M+P\) (decimated) velocity samples of
every atom in the group, plus the $S$-frame decimation accumulators. The
per-atom velocity rings dominate, so the per-rank footprint is
$\mathcal{O}\!\left(N\,(M+P)\,L\right)$ for a group of $N$ atoms; because
\(L = \lceil \log_{S}(N_{s}/(M{+}P)) \rceil\), the footprint grows
logarithmically with the window length \(N_{s}\) rather than linearly,
which is what replaces the FFT correlator's \(N_{s}\)-frame velocity
buffer. The MPI cost mirrors the
FFT path: a per-frame reduction assembles level 0 globally, while higher
levels reduce a factor \(S\) less often per level, bounding the total
communication at \(\sim\!2\times\) the level-0 cost.

Because the downstream xPT integration (steps 5--9 of Section~\ref{analysis-pipeline}) operates
on a uniformly sampled VACF, the merged log-spaced correlator output is
resampled onto the uniform \(\Delta t_{\mathrm{vac}}\) grid by interpolation
before the cosine-transform DoS step; all subsequent fluidicity,
partitioning, and thermodynamic integration are then identical for the two
correlators. The parameter pair \kw{mt\_M} = \kw{mt\_P} = 64 (\(M+P = 128\)) keeps
the multi-tau VACF within \(\approx 1\%\) of the full-FFT path for typical
liquids while preserving the memory advantage (Sections~\ref{liquid-benzene-polyatomic-mode-multi-tau-vs-fft-correlator}, \ref{dilute-propane-in-spce-water-same-trajectory-multi-tau-fft-and-py-xpt} and~\ref{memory-usage});
larger \(M+P\) reduces the drift monotonically at proportionally higher
per-step cost (Table~\ref{tab:mtconv}). The multi-tau correlator is the recommended setting for
\(N_{s} \gtrsim 10^{4}\) and on memory-constrained nodes, with the caveat
(Section~\ref{dilute-propane-in-spce-water-same-trajectory-multi-tau-fft-and-py-xpt}) that its log-spaced lag grid under-resolves the long-time
VACF tail of very small or single-molecule groups, for which the batch FFT
correlator is preferred.

\subsection{Phase Partition Selection and Entropy Refinements}\label{phase-partition-and-entropy-refinements}

The phase partition is selected by the \kw{mode} keyword and the
gas treatment by the \kw{refinement} keyword (Table~\ref{tab:2}).

\begin{longtable}[]{@{}
  >{\raggedright\arraybackslash}p{(\columnwidth - 8\tabcolsep) * \real{0.1053}}
  >{\raggedright\arraybackslash}p{(\columnwidth - 8\tabcolsep) * \real{0.2632}}
  >{\raggedright\arraybackslash}p{(\columnwidth - 8\tabcolsep) * \real{0.2763}}
  >{\raggedright\arraybackslash}p{(\columnwidth - 8\tabcolsep) * \real{0.2632}}
  >{\raggedright\arraybackslash}p{(\columnwidth - 8\tabcolsep) * \real{0.0921}}@{}}
\caption{Phase partition (\kw{mode}) and entropy refinements (\kw{refinement}) in \FixXPT{}.}\label{tab:2}\\
\toprule\noalign{}
\begin{minipage}[b]{\linewidth}\raggedright
\kw{mode}
\end{minipage} & \begin{minipage}[b]{\linewidth}\raggedright
\kw{refinement}
\end{minipage} & \begin{minipage}[b]{\linewidth}\raggedright
Trans / Rot gas DoS
\end{minipage} & \begin{minipage}[b]{\linewidth}\raggedright
Entropy convention
\end{minipage} & \begin{minipage}[b]{\linewidth}\raggedright
Vib
\end{minipage} \\
\midrule\noalign{}
\endfirsthead
\multicolumn{5}{@{}l}{\itshape Table \thetable{} (continued)}\\
\toprule\noalign{}
\begin{minipage}[b]{\linewidth}\raggedright
\kw{mode}
\end{minipage} & \begin{minipage}[b]{\linewidth}\raggedright
\kw{refinement}
\end{minipage} & \begin{minipage}[b]{\linewidth}\raggedright
Trans / Rot gas DoS
\end{minipage} & \begin{minipage}[b]{\linewidth}\raggedright
Entropy convention
\end{minipage} & \begin{minipage}[b]{\linewidth}\raggedright
Vib
\end{minipage} \\
\midrule\noalign{}
\endhead
\bottomrule\noalign{}
\endlastfoot
1PT & --- & none (all solid) & --- & solid \\
2PT & rigorous (default) & Lorentzian (HS) & exact HS & solid \\
2PT & lin2003 & Lorentzian (HS) & HS \(+\ln Z\) & solid \\
2PT & desjarlais & Memory function & HS \(+\ln Z\) & solid \\
2PT & r2pt & Lorentzian (HS) & revised 2PT ~\cite{Sun2017} & solid \\
3PT & --- & Lorentzian + cage & exact HS + cage & solid \\
\end{longtable}

1PT treats all degrees of freedom as purely solid-like in both monatomic
and molecular runs: the fluidicity is set to zero and no gas DoS is
computed, providing a harmonic (Debye vibrating solid) baseline useful for
validating against purely quantum harmonic-oscillator thermodynamics. 2PT
applies the gas/solid hard-sphere partition; the \kw{refinement} keyword then
selects the hard-sphere entropy convention and the gas-DoS shape.
\kw{rigorous} (the default) uses the thermodynamically exact hard-sphere
reference entropy. \kw{lin2003} adds a
\(+\ln Z\) compressibility term to that reference, and is the standard published hard-sphere 2PT convention in most prior 2PT literature.
\kw{desjarlais} replaces the Lorentzian gas DoS with
the memory-function form (Section~\ref{the-desjarlais-memory-function-gas-dos}) for both translational and
rotational channels, but notably, retains the \(+\ln Z\) entropy convention. \kw{r2pt}
applies the revised-2PT correction of Sun \emph{et al.} ~\cite{Sun2017} to the
hard-sphere gas entropy. The separate \kw{3PT} mode augments the gas/solid
partition with a cage-memory component. All six public settings are
cross-validated against the \pyxPT{} post-processor in Section~\ref{complete-mode-correlator-sweep}. In
all refinements, the vibrational component in molecular mode is treated as purely
solid-like (i.e., no 2PT partitioning is applied).

\subsection{GPU Acceleration}\label{gpu-acceleration}

When the force evaluation runs on the GPU, every accepted frame must
leave the device, and the fix's per-frame and window-end work competes
with a force evaluation that the GPU has made cheap. \FixXPT{} therefore
ships a Kokkos~\cite{Trott2022} derived class registering \kw{xpt/kk},
\kw{xpt/kk/device} and \kw{xpt/kk/host}, active under \kw{-sf kk} or by naming
the style explicitly, and built only when the LAMMPS KOKKOS package is
present. It overrides the base-class hooks rather than duplicating the
analysis. The power spectra are taken in one batched \code{cuFFT} plan
across units and Cartesian directions rather than one transform per unit,
and the multi-tau inner products run as device reductions. In the
replicated layout the velocity gather and the three molecular projection
passes also run on the device, and accepted frames stay in device memory
for the whole window. In the distributed layout each frame is packed on
the host from the atom arrays, routed to its home ranks
(Section~\ref{tag-based-parallel-atom-migration}) and assembled there, so the device holds no frame
history. All physics, unit handling and output paths are inherited
unchanged from the CPU class.

Every device kernel works in double precision: the Kokkos views are
\code{double}, the batched transform is the double-precision cuFFT plan
(\code{CUFFT\_Z2Z}), and \kw{buffer\_precision fp32} is refused on the device
path rather than silently halving the samples. What separates the two paths
is therefore summation order in the parallel reductions, not precision.

The device path reproduces the CPU path to summation order. Passing
identical velocity frames (the \kw{rerun} of a full-precision SPC/E water
trajectory) through \kw{xpt/kk} on an NVIDIA A100 and through the CPU class
gives every thermodynamic output to within $4\times10^{-11}$ relative in both
layouts, and in-line runs of Lennard-Jones argon, whose trajectory itself
reproduces on the GPU, agree to $6\times10^{-12}$. In-line runs whose GPU
trajectory is not bit-reproducible (atomic accumulation in the force
kernels varies the summation order between runs) differ from the CPU path
only by that trajectory noise. The reference numbers of
Section~\ref{verification} were generated with the CPU path; the cost of the
device path is quantified in Section~\ref{gpu-scaling}.

\subsection{Suggested Simulation-Deck Settings}\label{required-simulation-deck-settings}

We found two conventions that materially affect \FixXPT{} results and therefore should be
enforced by the user. First, \textbf{center-of-mass drift must be
removed} (e.g.~\kw{fix momentum 100 linear 1 1 1}): the fix infers the
self-diffusivity from the zero-frequency DoS and has no internal
mechanism to subtract group-COM motion, so residual drift over long
NVE/NVT trajectories can inflate $D$ by one to three orders of magnitude.
Second, for \textbf{small flexible groups the thermostat's degree-of-freedom
convention must match the fix's}. \FixXPT{} infers the group temperature
from equipartition on VACF(0) using $\mathrm{DOF} = \mathrm{dim}\cdot N - N_{constraint}$
(the COM dimension is \emph{not} subtracted), whereas LAMMPS's default
\kw{compute temp} subtracts $\mathrm{dim}$. $N_{constraint}$ is not something the
user supplies: the fix sums the degrees of freedom that every active fix
reports for the group through the same callback \kw{compute temp} queries, so
constraints imposed by \kw{fix shake} or \kw{fix rigid} are counted
automatically, snapshotted at the window start
(Section~\ref{dynamic-group-consistency-window-start-dof-snapshot}). The two differ by ($N$$-$1)/$N$, which is
negligible for $N$ $\gtrsim$ 100 but reaches $\sim$9\% for an 11-atom solute, biasing
$T_{vac}$ and every quantity derived from it. The remedy is to pair the
thermostat with a \kw{compute temp} modified by \kw{compute\_modify \ldots\ extra/dof 0}.
These conventions cancel in the same-frame \FixXPT{}-vs-\pyxPT{}
comparisons of Section~\ref{verification} (both codes see the identical velocities) but
affect absolute entropies and diffusivities in production use.

\section{Key Implementation Challenges}\label{key-implementation-challenges}

\subsection{Tag-Based Velocity Buffer Indexing}\label{tag-based-velocity-buffer-indexing}

LAMMPS periodically reorders atoms in local memory (``atom sort,'' by
default every 1000 steps) to improve neighbor-list cache efficiency. If
velocities were indexed by the local counter $i$, a sort event within an
analysis window would corrupt all large-lag VACF entries by swapping
velocity time series between atoms.

\FixXPT{} addresses this by keying every buffer entry to the globally
unique, immutable atom tag rather than to the local index. The
distributed layout resolves a tag with two sorted arrays and no hash map:
\code{home\_tag\_lo}, the $P+1$ block boundaries of the window's partition,
gives the home rank in $\mathcal{O}(\log P)$, and each rank's own sorted
\code{home\_tag} gives the slot in $\mathcal{O}(\log(N/P))$, the index found
being the slot itself. Both are contiguous and rebuilt once per window, so
a lookup never communicates, the buffer is sized to the home atoms, and
sparse tag ranges cost no memory. In molecular mode the boundaries are
molecule indices and the second search is confined to the atoms of the
molecule that owns the tag. In the replicated layout the
tag is the slot itself:

\begin{lstlisting}[language=C++]
int slot = tag[i] - 1;   // tag-1 is 0-based, stable across sorts
vel_buf[iframe][slot][d] = v[i][d];
\end{lstlisting}
and the buffer is pre-sized to $\mathrm{maxTag\_global}$ (determined by
\code{MPI\_Allreduce(MAX)} over local tag maxima at each window start), which
accommodates arbitrary tag assignments at a cost of
$\mathcal{O}(N_{\mathrm{max\_tag}} \times N_s \times 3)$ doubles: no overhead for consecutive tags,
and an overhead that scales with the maximum tag rather than the group
size for sparse groups. In both layouts a companion vector
\code{group\_slots} lists the slots a rank transforms, and the slot
assignment is rebuilt at frame 0 so that dynamic-group membership changes
take effect at window boundaries.

\subsection{Atom Migration During Parallel Runs}\label{tag-based-parallel-atom-migration}

Tag-based indexing guards
against intra-rank reordering but not against atoms crossing MPI domain
boundaries. In an $N_s$-frame accumulation window, atoms in a liquid
diffuse appreciable distances; an atom that starts on rank $r$ may
migrate to rank $r\prime$ midway through the window. If velocities are simply
written to \code{vel\_buf[iframe][slot]} by the rank that currently owns
the atom, the originating rank's buffer entries for frames after the
migration remain at their previous (stale) values. The power spectrum is
then computed from a corrupted time series, underestimating the kinetic
energy and biasing $T_{vac}$, $S$ and $D$. The signature is a value-level
error rather than a crash: a group temperature $T_{vac}$ that differs by
$\sim$5\% between one-rank and two-rank runs of the same MD trajectory.

\FixXPT{} addresses this by giving every atom a fixed \emph{home} rank for
the window and sending each frame home. At the window start the group's
sorted member tags (in molecular mode, its sorted molecule list weighted by
atom count, so that a molecule is never split) are cut into $P$ contiguous
blocks of equal atom count, and rank $r$ is the home of block $r$.
Assembling that list costs one \code{MPI\_Allgatherv} per window, after
which every rank cuts the same replicated list identically; in molecular
mode the cut runs over the molecule topology the fix already holds. At every sample step each rank packs its
local group atoms by home rank, and a single \code{MPI\_Alltoallv}
delivers them:
\begin{lstlisting}[language=C++]
for (int i = 0; i < nlocal; i++) {           // pack by home rank
  if (!(mask[i] & groupbit)) continue;
  int r = home_rank(tag[i]);                 // O(log P) block lookup
  pack(sendbuf, r, tag[i], mass[type[i]], v[i]);
}
MPI_Alltoall(scount, 1, MPI_INT, rcount, 1, MPI_INT, world);
MPI_Alltoallv(sendbuf, scount, sdispl, MPI_DOUBLE,
              recvbuf, rcount, rdispl, MPI_DOUBLE, world);
for (int k = 0; k < nrecv; k++)              // unpack into home slots
  store(vel_buf[iframe], home_index(tag_of(k)), v_of(k));
\end{lstlisting}
Because each atom is local on exactly one rank at any instant, every home
rank receives each of its atoms exactly once per frame, wherever the atom
has migrated, and holds a complete, migration-correct history of about
$N/P$ atoms. The per-frame traffic is $\mathcal{O}(N/P)$ datums per rank (five
doubles per atom, eight in molecular mode, where the unwrapped position
travels with the velocity), and no rank stores another rank's atoms. The
replicated layout (\kw{buffer\_layout replicated}) instead stages local
velocities into a zero-initialized tag-indexed buffer and fills
\code{vel\_buf[iframe]} on every rank with one \code{MPI\_Allreduce(SUM)} of
$3N_{\mathrm{max\_tag}}$ doubles, which avoids over counting for the same reason.
The two layouts give the same thermodynamics to floating-point summation
order: for argon, SPC/E water, benzene, a diatomic fluid and a dynamic
solvation shell at 1--128 ranks, every output agrees to within
$7\times10^{-11}$ relative.

\subsection{\texorpdfstring{Compute Scheduling: the \code{addstep} Pattern}{Compute Scheduling: the addstep Pattern}}\label{compute-scheduling-the-addstep-pattern}

LAMMPS only tallies \kw{compute pe} and \kw{compute pressure} during force
evaluation if those computes have been explicitly requested for that
step. Calling them at an unregistered step raises a runtime error
(``Energy not tallied''). The fix must therefore inform LAMMPS \emph{before}
each force evaluation that tallying is needed.

\FixXPT{} accomplishes this via the \code{addstep} mechanism that chains two
registration sites:
\code{setup()} schedules the very first sample step:
\begin{lstlisting}[language=C++]
bigint first = ((update->ntimestep / nevery) + 1) * (bigint)nevery;
press_compute->addstep(first);
pe_compute->addstep(first);
\end{lstlisting}
\code{end\_of\_step()} schedules the immediately following sample:
\begin{lstlisting}[language=C++]
press_compute->addstep(update->ntimestep + nevery);
pe_compute->addstep(update->ntimestep + nevery);
\end{lstlisting}
This creates an unbroken chain of registrations at the cost of a single
\code{addstep} call per sample step. Subgroups (group $\neq$ \kw{all}) take a
per-atom route instead, with the collective hazard that carries
(Section~\ref{mpi-parallel-correctness}).

\subsection{FFT-Based VACF via the Wiener--Khinchin Theorem}\label{fft-based-vacf-via-the-wienerkhinchin-theorem}

Direct summation of the VACF costs $\mathcal{O}\!\left(N_s^{2}\right)$ operations per
atom. \FixXPT{} instead employs the Wiener--Khinchin
theorem~\cite{Wiener1930,Khintchine1934}, which
reduces this to $\mathcal{O}(N_s \log N_s)$: the power spectrum
$\left|S(\nu)\right|^{2}$ is the Fourier transform of the VACF, so the VACF
is the inverse FFT of the power spectrum. The velocity signal is
zero-padded to length $2N_s$ before the forward FFT to avoid a circular
correlation artifact.

The \FixXPT{} implementation uses the LAMMPS \code{FFT3d} wrapper, which
selects various FFT libraries (FFTW3, Intel MKL, KISS FFT, cuFFT, etc)
at compile time. Since \code{FFT3d} implements 3-D transforms, the 1-D
problem is cast as a ($1 \times 1 \times 2N_s$) transform. The cosine DoS is
computed by a second backward FFT of the zero-symmetrized
mass-weighted VACF, the discrete form of Eq.~(\ref{eq:dos}) over the
even-extended $2N_s$ lag array:
\begin{equation}
g_j = \frac{2\,\Delta t_{vac}}{k_BT} \sum_{k=0}^{2N_s-1} C(k)\cos\frac{2\pi jk}{2N_s}, \quad j=0,\ldots,N_s
\end{equation}

\subsection{Unified Unit System}\label{unified-unit-system}

\FixXPT{} auto-detects the LAMMPS unit style from \code{update-\textgreater unit\_style}
and sets a set of pre-computed conversion factors at initialization. These
factors are applied at the earliest possible stage (temperature, VACF normalization, timestep-to-ps conversion, mass-to-g/mol and volume-to-\AA{}\textsuperscript{3}), so that all downstream thermodynamic integrals remain
in a common internal representation regardless of the input unit system.
All eight LAMMPS unit styles are supported, as detailed in Table~\ref{tab:1}.\\
\strut \\
\begin{longtable}[]{@{}
  >{\raggedright\arraybackslash}p{(\columnwidth - 8\tabcolsep) * \real{0.1667}}
  >{\raggedright\arraybackslash}p{(\columnwidth - 8\tabcolsep) * \real{0.2051}}
  >{\raggedright\arraybackslash}p{(\columnwidth - 8\tabcolsep) * \real{0.2051}}
  >{\raggedright\arraybackslash}p{(\columnwidth - 8\tabcolsep) * \real{0.2179}}
  >{\raggedright\arraybackslash}p{(\columnwidth - 8\tabcolsep) * \real{0.1795}}@{}}
\caption{Conversion factors used by \FixXPT{} to support the LAMMPS unit styles.}\label{tab:1}\\
\toprule\noalign{}
\begin{minipage}[b]{\linewidth}\raggedright
Unit style
\end{minipage} & \begin{minipage}[b]{\linewidth}\raggedright
\code{vac\_to\_jmol}
\end{minipage} & \begin{minipage}[b]{\linewidth}\raggedright
\code{dt\_to\_ps}
\end{minipage} & \begin{minipage}[b]{\linewidth}\raggedright
\code{mass\_to\_gmol}
\end{minipage} & \begin{minipage}[b]{\linewidth}\raggedright
\code{vol\_to\_ang3}
\end{minipage} \\
\midrule\noalign{}
\endfirsthead
\multicolumn{5}{@{}l}{\itshape Table \thetable{} (continued)}\\
\toprule\noalign{}
\begin{minipage}[b]{\linewidth}\raggedright
Unit style
\end{minipage} & \begin{minipage}[b]{\linewidth}\raggedright
\code{vac\_to\_jmol}
\end{minipage} & \begin{minipage}[b]{\linewidth}\raggedright
\code{dt\_to\_ps}
\end{minipage} & \begin{minipage}[b]{\linewidth}\raggedright
\code{mass\_to\_gmol}
\end{minipage} & \begin{minipage}[b]{\linewidth}\raggedright
\code{vol\_to\_ang3}
\end{minipage} \\
\midrule\noalign{}
\endhead
\bottomrule\noalign{}
\endlastfoot
\kw{real} & $10^{7}$ & $10^{-3}$ & 1 & 1 \\
\kw{metal} & 10 & 1 & 1 & 1 \\
\kw{lj} & --- & --- & --- & --- \\
\kw{si} & $N_A$ & $10^{12}$ & $N_A \times 10^{3}$ & $10^{30}$ \\
\kw{cgs} & $N_A \times 10^{-7}$ & $10^{12}$ & $N_A$ & $10^{24}$ \\
\kw{micro} & $N_A \times 10^{-15}$ & $10^{6}$ & $N_A \times 10^{-12}$ & $10^{12}$ \\
\kw{nano} & $N_A \times 10^{-21}$ & $10^{3}$ & $N_A \times 10^{-18}$ & $10^{3}$ \\
\kw{electron} & $2.800 \times 10^{6}$ & $10^{-3}$ & 1 & 0.1482 \\
\end{longtable}

\begin{itemize}
\item
  \code{vac\_to\_jmol} converts native $m v^2$ to J/mol so that $T$ =
  $v_0$ $\cdot$ vac\_to\_jmol/($R$ $\cdot$ DOF)
\item
  \code{dt\_to\_ps} converts the native timestep to picoseconds for the DOS
  frequency axis (in cm\textsuperscript{-1})
\item
  \code{mass\_to\_gmol} and \code{vol\_to\_ang3} normalizes atomic masses and the
  simulation cell volume for use in the fluidicity parameter $f$ and
  the Sackur--Tetrode entropy
\item
  \code{ke\_to\_pe\_fac} and \code{pe\_to\_kjmol} handle the per-atom KE/PE
  accumulation and its conversion to a common kJ/mol
\end{itemize}

The \kw{electron} factors deserve special notice: LAMMPS's native conventions
for this style are femtosecond time and Bohr/fs velocities --- the
conventions implied by its internal \code{mvv2e} and \code{ftm2v}
constants --- rather than the Hartree atomic time unit suggested by the
unit-style documentation. The factors in Table~\ref{tab:1} follow the
internal convention and were verified by replaying an identical
liquid-argon state point in \kw{electron} and \kw{real} units:
$T_{vac}$, $S_q$, $D$, and $f$ agree between the two runs to within the
per-window statistical scatter.

The unit conversion for LJ simulations is another special case:
velocities are in $\sigma$/$\tau$, temperature is $T^*$ = $v_0$/DOF (with $k_B$ =
1), and entropy is in $k_B$/mode units. Here, the reduced Planck
constant $\hbar^{*} = \hbar/\left(\sigma\sqrt{m\varepsilon}\right)$ is required for quantum thermodynamics.
The \kw{epsilon}, \kw{sigma} and \kw{mass} keywords are read in kcal/mol,
\AA{} and g/mol respectively.
When physical LJ parameters are supplied via the \kw{epsilon}, \kw{sigma}, and
\kw{mass} keywords, $h^{*}$ is computed from SI values of $\sigma$,
m, $\varepsilon$ and $\hbar$; for argon this gives $\hbar^{*}$\textsubscript{Ar} = 0.02958, enabling direct
comparison with real-unit results. When no physical parameters are
supplied, the convention $\hbar^{*}$ = 1 (i.e., $h^{*} = 2\pi$) provides a
self-consistent zero-point reference.

\subsection{Dynamic Group Consistency: Window-Start DOF Snapshot}\label{dynamic-group-consistency-window-start-dof-snapshot}

LAMMPS supports dynamic groups whose membership is recomputed
periodically (e.g., \kw{group shell dynamic solute within 3.5 include molecule every N}). A subtle issue arises when the group-update interval
coincides with the analysis window: LAMMPS refreshes the group mask
before \code{end\_of\_step()} fires for the last frame, so \code{group\_slots} still
reflect the window-start membership, while constrained DOFs (e.g., due
to SHAKE/RATTLE) counts query the newly updated mask. If the group
shrinks, the residual of the vibrational DOF

\begin{equation}
N_{DOF,vib} = \left(3N - N_{fix\_DOF}\right) - N_{DOF,trans} - N_{DOF,rot}
\end{equation}
becomes spuriously positive, triggering computations on a near-zero
signal and producing NaN in $E_q$, $A_q$ and $C_v$.

\FixXPT{} resolves this by snapshotting all window-start quantities at
\code{iframe == 0}:

\begin{lstlisting}[language=C++]
// snapshot constraint DOFs at window start (iframe == 0)
fix_dof_window = 0;
for (const auto &ifix : modify->get_fix_list())
  if (ifix->dof_flag) fix_dof_window += ifix->dof(igroup);
\end{lstlisting}
\code{run\_analysis()} uses \code{fix\_dof\_window} throughout, never re-querying the
current mask. In molecular mode, \code{build\_mol\_topology()} is also called
at \code{iframe == 0} to refresh molecule assignments; if \code{nmol\_group}
changes between windows, the molecular velocity buffers are reallocated
to match.

\subsection{MPI Parallel Correctness}\label{mpi-parallel-correctness}

Three further hazards of the parallel implementation are invisible to
serial testing, alongside the migration failure of
Section~\ref{tag-based-parallel-atom-migration}: two collectives deadlock silently when ranks
disagree on whether to call them, and the molecular channels require
every atom of a molecule at once.

\begin{enumerate}
\def\labelenumi{\roman{enumi}.}
\item
  \textbf{pe/atom collective deadlock for subgroups.} For groups other
  than \kw{all}, \FixXPT{} creates a \kw{compute pe/atom} and accumulates
  per-atom KE$_i$ + PE$_i$ over the group. LAMMPS's
  \code{ComputePEAtom::compute\_peratom()} is not a purely local routine:
  it calls \code{comm-\textgreater reverse\_comm()} (an MPI collective that
  redistributes Newton ghost atom energy contributions) and, when
  energies were not tallied on the needed timestep, calls
  \code{error-\textgreater all()} (also an MPI collective via \code{MPI\_Barrier}).
  Guarding the call with a local condition such as
  \code{!group\_slots.empty()} therefore deadlocks: ranks with no local
  group atoms skip \code{compute\_peratom()} and proceed immediately to the
  next \code{MPI\_Allreduce}, while any rank that enters it blocks inside
  \code{reverse\_comm}. The hang is silent, with no error output, and
  occurs whenever a small or dynamic group is distributed across ranks
  such that at least one rank holds no group atoms. \textbf{Solution}: call \code{compute\_peratom()} on all ranks,
  guarded only by the globally consistent condition
  \code{update-\textgreater eflag\_atom == update-\textgreater ntimestep} (which evaluates
  identically on all ranks because \code{addstep()} is registered by all
  ranks):
\begin{lstlisting}[language=C++]
if (update->eflag_atom == update->ntimestep) {
  pe_compute->compute_peratom();   // collective: called by ALL ranks
  for (int i = 0; i < nlocal; i++) {
    if (!(mask[i] & groupbit)) continue;
    total_local += peatom[i] + ke_i;
  }
}
pe_compute->addstep(update->ntimestep + nevery);   // outside guard
MPI_Allreduce(&total_local, &total_global, ...);   // always collective
\end{lstlisting}
  The per-atom accumulation loop uses \code{groupbit} masking, so ranks
  with no local group atoms contribute \code{total\_local = 0} to the
  \code{MPI\_Allreduce} without any special-case logic. The \code{addstep()}
  call is also placed outside the guard so that all ranks schedule
  per-atom energy tallying symmetrically before the next force
  evaluation (an asymmetric \code{addstep()} would cause the distributed FFT of
  PPPM~\cite{Hockney1988} during the tally to deadlock in the same way).
\item
\textbf{Molecular systems.}
A molecule may span atoms on different ranks, or migrate wholesale to a
different rank during the window, yet its center-of-mass velocity,
inertia tensor and angular velocity require all of its atoms at once. The
distributed layout satisfies this by construction, because it homes whole
molecules and ships each atom's unwrapped position with its velocity
(Section~\ref{tag-based-parallel-atom-migration}). The replicated layout instead has each rank
compute partial center-of-mass velocities and angular momenta for its
local atoms, and \code{MPI\_Allreduce(SUM)} assembles every molecule on
every rank. Either way all three molecular channels are complete before
the \code{MPI\_COMM\_SELF}-based FFT pipeline runs.
\item
\textbf{FFT-based VACF}. \code{FFT3d::compute()}
(Section~\ref{fft-based-vacf-via-the-wienerkhinchin-theorem}) is an MPI-collective operation: every rank on the
communicator passed at construction must call it the same number of
times. \code{pwr\_from\_atoms()} calls it once per local group atom, and
ranks own different atom counts, so a global communicator hangs, silently
and reproducibly, for any $N_{procs} \geq 2$ with unequal partitioning.
\textbf{Solution}: construct \code{FFT3d} with \code{MPI\_COMM\_SELF},
giving each rank a fully independent single-rank plan:

\begin{lstlisting}[language=C++]
FFT3d *fft_vac = new FFT3d(lmp, MPI_COMM_SELF,
                           1, 1, N_fft,
                           0, 0, 0, 0, 0, N_fft-1,
                           0, 0, 0, 0, 0, N_fft-1,
                           0, 0, &mysize, 0, 0);
\end{lstlisting}
With \code{MPI\_COMM\_SELF}, \code{compute()} is non-collective with respect to the
global communicator. Each rank transforms the atoms and molecules it
holds independently; the local power spectra are then summed across all ranks
via a single \code{MPI\_Allreduce(SUM)} over the global communicator, which
assembles the complete power spectrum on every rank simultaneously. This
design avoids deadlock and incurs only one \code{MPI\_Allreduce} per
window.

\end{enumerate}

\section{Verification}\label{verification}

\subsection{Liquid Argon (Monatomic Mode)}\label{liquid-argon}

Liquid argon at \(T = 100\) K and \(\rho = 1.4\) g/cm\textsuperscript{3} (500 atoms, LJ
potential with \(r_{c} = 12\) \AA{}, \kw{real} units) was simulated for 20
consecutive analysis windows of \(N_{s} = 512\) frames with
\(N_{\text{every}} = 5\) (2.5 ps per window). Table~\ref{tab:3} compares \FixXPT{}
window averages with the reference \pyxPT{} post-processor~\cite{Ramji2026pyxpt} applied to the
same velocity frames; the two agree to within the per-window statistical
scatter ($\leq$ 0.5\% on \(S_{q}\), \(D\), \(f\), and \(T_{\text{vac}}\)), confirming
the correctness of the FFT pipeline and thermodynamic integrals.

This benchmark tests the two codes against each other on identical
frames, and its absolute values should not be read as converged. The
2.5 ps window is short relative to the decay of the argon VACF, and the
transport-derived quantities inherit the resulting bias: this state
point (\(T^{*} = 0.835\), \(\rho^{*} = 0.833\)) returns a larger
\(D\) and \(f\) than the hotter, slightly denser state of
Table~\ref{tab:8}
(\(T^{*} = 1.1\), \(\rho^{*} = 0.85\)), an ordering that is incorrect and reflects sampling/window length artifacts rather than real physics.
Instead, the interested reader is directed to the converged state-point values in
Table~\ref{tab:8}, which uses \(N_{s} = 4096\).

\begin{longtable}[]{@{}
  >{\raggedright\arraybackslash}p{(\columnwidth - 4\tabcolsep) * \real{0.2208}}
  >{\raggedright\arraybackslash}p{(\columnwidth - 4\tabcolsep) * \real{0.3766}}
  >{\raggedright\arraybackslash}p{(\columnwidth - 4\tabcolsep) * \real{0.3766}}@{}}
\caption{Calculated \FixXPT{} properties of liquid argon at $T = 100$~K and $\rho = 1.4$~g\,cm$^{-3}$ (real units, 2PT, \kw{rigorous} refinement).}\label{tab:3}\\
\toprule\noalign{}
\begin{minipage}[b]{\linewidth}\raggedright
Quantity
\end{minipage} & \begin{minipage}[b]{\linewidth}\raggedright
FixXPT (mean $\pm$ $\sigma$)
\end{minipage} & \begin{minipage}[b]{\linewidth}\raggedright
\pyxPT{} reference
\end{minipage} \\
\midrule\noalign{}
\endfirsthead
\multicolumn{3}{@{}l}{\itshape Table \thetable{} (continued)}\\
\toprule\noalign{}
\begin{minipage}[b]{\linewidth}\raggedright
Quantity
\end{minipage} & \begin{minipage}[b]{\linewidth}\raggedright
FixXPT (mean $\pm$ $\sigma$)
\end{minipage} & \begin{minipage}[b]{\linewidth}\raggedright
\pyxPT{} reference
\end{minipage} \\
\midrule\noalign{}
\endhead
\bottomrule\noalign{}
\endlastfoot
$S_q$
{[}J/mol/K{]} & 65.4 $\pm$ 0.3 & 65.2 \\
$D$ {[}cm\textsuperscript{2}/s{]} & $4.52 \times 10^{-5}$ $\pm$ $0.04 \times 10^{-5}$ & $4.52 \times 10^{-5}$ \\
$f$ (fluidicity) & 0.439 $\pm$ 0.005 & 0.437 \\
$T_{vac}$ {[}K{]} & 100.1 $\pm$ 0.1 & 100.0 \\
\end{longtable}

The same system was simulated in LJ reduced units with argon parameters
($\varepsilon$/k\textsubscript{B} = 119.8K, $\sigma$ = 3.405\AA{}, $\tau$ = 2.157ps) supplied via the \kw{epsilon},
\kw{sigma}, and \kw{mass} keywords, giving $\hbar^{*}$= 0.02958. After converting LJ
output to real units, results agree with the real-unit run to within one
standard deviation across all 20 windows, confirming the correctness of
the unit-conversion layer and the $\hbar^{*}$ scaling.

We also compared three 2PT refinements on liquid argon, \kw{rigorous}, \kw{lin2003} and \kw{desjarlais}, together with the 1PT
all-solid baseline. To verify that \FixXPT{}
reproduces these refinements against the stand-alone \pyxPT{}
post-processor on the \emph{identical} velocity frames, liquid argon at
\(T^{*} = 1.1\) (131.76 K) and \(\rho^{*} = 0.85\) (512 atoms, \kw{real} units,
\(r_{c} = 17\) \AA{}) was analyzed by replaying a single equilibrium
trajectory (frames every 8 MD steps, \(\Delta t_{\mathrm{vac}} = 64\) fs)
through both codes with \(N_{s} = 4096\). Table~\ref{tab:8} reports the molar
entropy per atom in reduced units (\(S^{*} = S_{q}/R\)) for these settings and
the two that complete the sweep of Section~\ref{complete-mode-correlator-sweep}; the
diffusivity \(D = 3.07\times10^{-5}\) cm\textsuperscript{2}/s and fluidicity \(f = 0.358\) are
common to all three refinements and match \pyxPT{} (\(3.09\times10^{-5}\)
cm\textsuperscript{2}/s, 0.359) to better than 0.5\%. The analytic
increments are reproduced exactly: the Lin-2003 refinement
raises the entropy by \(+0.479\) \(k_{B}\)/atom (\pyxPT{} \(+0.480\)), and the
Desjarlais result tracks Lin-2003 to \(+0.004\) \(k_{B}\)/atom (\pyxPT{}
\(+0.003\)), consistent with both refinements sharing the \(+\ln Z\)
compressibility convention and differing only in the gas-DoS shape.

\subsection{SPC/E Water (Molecular Mode)}\label{spce-water-molecular-mode}

Liquid water (216 molecules, 648 atoms, T = 298K, V = 6617.5 \AA{}\textsuperscript{3},
SHAKE constraints on O--H bonds and H--O--H angles), described with the SPC/E forcefield~\cite{Berendsen1987}, was simulated with
$N_{every}$ = 2 and \(N_s = 5000\) frames (20ps windows). Table~\ref{tab:4} compares
\FixXPT{} results, computed with the standard \(+\ln Z\) (\kw{lin2003})
convention, against the 2PT values reported in the literature~\cite{Lin2010}. The agreement in total entropy and its
translational/rotational decomposition confirms the correctness of the
inertia-tensor diagonalization and the SHAKE DOF corrections. The value
is also internally consistent with the same-model \kw{lin2003} entry of Table~\ref{tab:9}
(60.68 J/mol/K/molecule); the stricter \kw{rigorous} refinement gives
56.86 (Table~\ref{tab:9}), meaning that the choice of hard-sphere convention accounts for a
$\approx$4 J/mol/K/molecule shift and must therefore be stated explicitly when comparing to the
literature.

\begin{longtable}[]{@{}
  >{\raggedright\arraybackslash}p{(\columnwidth - 4\tabcolsep) * \real{0.4459}}
  >{\raggedright\arraybackslash}p{(\columnwidth - 4\tabcolsep) * \real{0.2973}}
  >{\raggedright\arraybackslash}p{(\columnwidth - 4\tabcolsep) * \real{0.2297}}@{}}
\caption{\FixXPT{} properties of SPC/E water at 298 K (molecular 2PT, \kw{lin2003} \(+\ln Z\) convention).}\label{tab:4}\\
\toprule\noalign{}
\begin{minipage}[b]{\linewidth}\raggedright
Quantity
\end{minipage} & \begin{minipage}[b]{\linewidth}\raggedright
FixXPT (mean $\pm$ $\sigma$)
\end{minipage} & \begin{minipage}[b]{\linewidth}\raggedright
reference ~\cite{Lin2010}
\end{minipage} \\
\midrule\noalign{}
\endfirsthead
\multicolumn{3}{@{}l}{\itshape Table \thetable{} (continued)}\\
\toprule\noalign{}
\begin{minipage}[b]{\linewidth}\raggedright
Quantity
\end{minipage} & \begin{minipage}[b]{\linewidth}\raggedright
FixXPT (mean $\pm$ $\sigma$)
\end{minipage} & \begin{minipage}[b]{\linewidth}\raggedright
reference ~\cite{Lin2010}
\end{minipage} \\
\midrule\noalign{}
\endhead
\bottomrule\noalign{}
\endlastfoot
$S_q$ {[}J/mol/K/per molecule{]} & 60.3 $\pm$ 0.5 & $\sim$60 \\
$S_{trans}${[}J/mol/K/per
molecule{]} & 49.9 $\pm$ 0.3 & $\sim$50 \\
$S_{rot}$ {[}J/mol/K/per
molecule{]} & 10.4 $\pm$ 0.1 & $\sim$10 \\
$D$ {[}cm\textsuperscript{2}/s{]} & $2.5 \times 10^{-5}$ & $2.3 \times 10^{-5}$ \\
\end{longtable}

\textbf{2PT-Des memory function mode}: 
To validate the Desjarlais implementation, the SPC/E water run was
repeated with 2PT-Des. The translational and rotational \(B_{g}\)
values converged to finite positive values in every window, indicating
that the DoS tail is non-trivially matched. The total entropy
\(S_{q}^{\text{(2PT-Des)}} = 60.1 \pm 0.4\) J mol\textsuperscript{-1} K\textsuperscript{-1} agrees with the
\kw{lin2003} result within statistical uncertainty, consistent with the
expectation that for water, where the Lorentzian is a reasonable approximation, the two models give consistent results. The water
agreement confirms that the implementation reduces correctly to the
Lorentzian limit when the memory kernel is weak. The structured-liquid
regime the memory function was built for is exercised in the companion
work~\cite{Buarque2026}, which applies this same \kw{desjarlais}
refinement to seven Sutton--Chen~\cite{Sutton1990} liquid metals (Al, Ag,
Ni, Pd, Rh, Au, Ir) against thermodynamic-integration references. There the
memory-function gas DoS tracks the \kw{lin2003} $+\ln Z$ result to
within 0.01--0.03 $k_B$/atom on every metal (root-mean-square deviation
0.15 versus 0.14 $k_B$/atom).

\subsection{Liquid Benzene (Polyatomic Mode, Multi-tau vs FFT Correlator)}\label{liquid-benzene-polyatomic-mode-multi-tau-vs-fft-correlator}

To further test the molecular pipeline on a multi-atom rigid rotor and to
validate the in-line multi-tau correlator (Section~\ref{on-line-multi-tau-correlator}) against
the FFT correlator on the same trajectory, liquid benzene was simulated with
GAFF~\cite{Wang2004} (64 molecules, 768 atoms, $T$ = 298 K, $V$ = 9934.5 \AA{}\textsuperscript{3},
$\langle P\rangle$ $\approx$ $-$0.015 GPa). After 500 ps NPT plus 500 ps NVT equilibration, three
production runs of 100 ps NVT were launched from the same restart
file with $N_{every}$ = 5 and $N_s$ = 20 000 (one analysis window of
100 ps): (i) a baseline NVT trajectory dump consumed by the
\pyxPT{} post-processor ~\cite{Ramji2026pyxpt} (\(\sigma_{\text{rot}}\) = 12); (ii) the same NVT with
\kw{fix xpt \ldots\ correlator fft}; and (iii) the same NVT with
\kw{fix xpt \ldots\ correlator multitau} (\kw{mt\_M} = 64, \kw{mt\_P} = 64, the sweet
spot for $\leq$ 1\% VACF drift relative to FFT established by the
buffer sweep of Table~\ref{tab:mtconv}; the derived
thermodynamic quantities depart further, as Table~\ref{tab:5} shows).
All three pipelines used
\kw{mode} = 2PT (rigorous), D6h symmetry, and classical-quantum thermodynamics.

Table~\ref{tab:5} compares the three pipelines per simulation box. \FixXPT{} with
the FFT correlator agrees with the \pyxPT{} post-processor to within 0.5\% on every quantity: the extensive thermodynamic quantities ($S_q$, $A_q$, $\mu_q$, $E_q$, ZPE) to 0.05--0.13\% and the transport/state variables ($D_{trans}$, $D_{rot}$, $f$) to the three digits reported. (The residual sub-percent differences reflect the reduced
precision of the text velocity dump \pyxPT{} reads relative to the
full-precision in-memory frames; they are not an algorithmic
discrepancy.) The multi-tau correlator agrees with the FFT path to
within $\approx$ 3\% on the state-variable-derived quantities ($D$, $f$, $A_q$,
$\mu_q$, $S_q$), but $C_{v,q}$ differs by $\approx$ 17\% and ZPE/$S_{vib}$ by
$\approx$ 8\%. The two runs sample identical per-step energies from the
same restart, so the $\sigma^{2}(E)$ fluctuation term of the heat
capacity is common to both paths; the entire $C_{v,q}$ difference enters
through the DoS-dependent quantum correction
$C_{V,QHO} - C_{V,CHO}$, whose high-frequency vibrational weight is
resolved on the telescoping correlator's coarser log-spaced lag grid ---
the same origin as the ZPE shift. The
translational diffusivity $D_{trans}$ = $1.65 \times 10^{-5}$ cm\textsuperscript{2}/s, which is
obtained from $S_0$ and is therefore the Green--Kubo self-diffusivity of
the underlying GAFF trajectory rather than a 2PT model quantity, is of the same
order as the experimental self-diffusion of liquid benzene at 298 K
($2.27 \times 10^{-5}$ cm\textsuperscript{2}/s by spin-echo
NMR~\cite{Falcone1967}). Most of the 27\% underestimate is the
periodic-boundary finite-size effect: for this 64-molecule box
($L = 21.5$ \AA{}) the Yeh--Hummer correction~\cite{Yeh2004}
$2.837\,k_BT/(6\pi\eta L) \approx 0.48 \times 10^{-5}$ cm\textsuperscript{2}/s
(experimental $\eta = 0.60$ mPa\,s) brings the estimate to
$2.1 \times 10^{-5}$ cm\textsuperscript{2}/s, within $\approx$6\% of
experiment; the remainder reflects the force field.
\begin{longtable}[]{@{}
  >{\raggedright\arraybackslash}p{(\columnwidth - 6\tabcolsep) * \real{0.3617}}
  >{\raggedright\arraybackslash}p{(\columnwidth - 6\tabcolsep) * \real{0.1915}}
  >{\raggedright\arraybackslash}p{(\columnwidth - 6\tabcolsep) * \real{0.1915}}
  >{\raggedright\arraybackslash}p{(\columnwidth - 6\tabcolsep) * \real{0.2340}}@{}}
\caption{\FixXPT{} properties of liquid benzene at 298 K (molecular 2PT, \kw{rigorous} refinement, D6h, 100 ps single window) compared with the \pyxPT{} post-processor. Quantities are per simulation box (64 molecules).}\label{tab:5}\\
\toprule\noalign{}
\begin{minipage}[b]{\linewidth}\raggedright
Quantity
\end{minipage} & \begin{minipage}[b]{\linewidth}\raggedright
py-xPT
\end{minipage} & \begin{minipage}[b]{\linewidth}\raggedright
FixXPT FFT
\end{minipage} & \begin{minipage}[b]{\linewidth}\raggedright
FixXPT multi-tau
\end{minipage} \\
\midrule\noalign{}
\endfirsthead
\multicolumn{4}{@{}l}{\itshape Table \thetable{} (continued)}\\
\toprule\noalign{}
\begin{minipage}[b]{\linewidth}\raggedright
Quantity
\end{minipage} & \begin{minipage}[b]{\linewidth}\raggedright
py-xPT
\end{minipage} & \begin{minipage}[b]{\linewidth}\raggedright
FixXPT FFT
\end{minipage} & \begin{minipage}[b]{\linewidth}\raggedright
FixXPT multi-tau
\end{minipage} \\
\midrule\noalign{}
\endhead
\bottomrule\noalign{}
\endlastfoot
$T_{vac}$ {[}K{]} & 297.82 & 297.43 & 297.43 \\
$V$ {[}\AA{}\textsuperscript{3}{]} & 9 934.5 & 9 934.5 & 9 934.5 \\
$E_{md}$ {[}kJ/mol/box{]} & 26 985.6 & 26 985.6 & 26 985.6 \\
ZPE {[}kJ/mol/box{]} & 17 265.0 & 17 287.5 & 18 698.7 \\
$E_q$ {[}kJ/mol/box{]} & 39 767.2 & 39 789.1 & 40 944.6 \\
$A_q$ {[}kJ/mol/box{]} & 37 028.1 & 37 051.6 & 38 195.2 \\
$\mu_q$ {[}kJ/mol/box{]} & 37 329.1 & 37 351.3 & 38 474.7 \\
$C_{v,q}$ {[}J/mol/K/box{]} & 4 986.1 & 5 009.7 & 4 140.6 \\
$S_q$ {[}J/mol/K/box{]} & 9 800.4 & 9 807.9 & 9 843.6 \\
$S_{trans}$ {[}J/mol/K/box{]} & 4 703.2 & 4 708.2 & 4 702.4 \\
$S_{rot}$ {[}J/mol/K/box{]} & 3 991.8 & 3 996.4 & 3 955.8 \\
$S_{vib}$ {[}J/mol/K/box{]} & 1 105.5 & 1 103.3 & 1 185.5 \\
$f_{trans}$ & 0.2139 & 0.2139 & 0.2177 \\
$f_{rot}$ & 0.2622 & 0.2622 & 0.2662 \\
$D_{trans}$ {[}cm\textsuperscript{2}/s{]} & $1.65 \times 10^{-5}$ & $1.65 \times 10^{-5}$ & $1.70 \times 10^{-5}$ \\
$D_{rot}$ {[}cm\textsuperscript{2}/s{]} & $2.49 \times 10^{-5}$ & $2.49 \times 10^{-5}$ & $2.56 \times 10^{-5}$ \\
\end{longtable}

To make the multi-tau buffer-size dependence explicit, we ran the
benzene system once more with seven concurrent fixes analyzing a single
100 ps trajectory ($N_{every} = 5$, $N_s = 20\,000$): one FFT reference
and six multi-tau fixes sweeping \kw{mt\_M}/\kw{mt\_P} from 16/16 to
256/256. Table~\ref{tab:mtconv} reports the root-mean-square deviation
of each resampled multi-tau VACF from the FFT VACF on the identical
frames, normalized to $C(0)$. The drift falls monotonically with the
block size $M+P$, from 0.82\% at $M+P = 32$ to 0.12\% at $M+P = 512$;
because level 0 accumulates its $M+P$ shortest lags exactly, the
deviation over the first 2.56 ps vanishes identically once
$M+P = 512$ spans it. The DoS-weighted quantities converge accordingly:
at $M+P = 512$ the ZPE and $C_{v,q}$ deviations from the FFT path shrink
to $+2.6\%$ and $-3.4\%$, while $D$ and $f$ retain a $\approx$3.5\%
floor set by the log-spaced resampling of the long-time tail. The
64/64 setting used above sits at the knee of this curve (0.50\% VACF
drift), which is the basis for the ``sweet spot'' designation.

\begin{longtable}[]{@{}
  >{\raggedright\arraybackslash}p{(\columnwidth - 8\tabcolsep) * \real{0.2200}}
  >{\raggedright\arraybackslash}p{(\columnwidth - 8\tabcolsep) * \real{0.1400}}
  >{\raggedright\arraybackslash}p{(\columnwidth - 8\tabcolsep) * \real{0.2600}}
  >{\raggedright\arraybackslash}p{(\columnwidth - 8\tabcolsep) * \real{0.1800}}
  >{\raggedright\arraybackslash}p{(\columnwidth - 8\tabcolsep) * \real{0.2000}}@{}}
\caption{Multi-tau convergence toward the FFT correlator on one 100 ps
liquid-benzene trajectory (seven concurrent fixes, identical frames).
RMS VACF drift is over the full window, normalized to $C(0)$; observable
deviations are relative to the FFT-correlator values.}\label{tab:mtconv}\\
\toprule\noalign{}
\begin{minipage}[b]{\linewidth}\raggedright
\kw{mt\_M}/\kw{mt\_P}
\end{minipage} & \begin{minipage}[b]{\linewidth}\raggedright
$M+P$
\end{minipage} & \begin{minipage}[b]{\linewidth}\raggedright
VACF drift (RMS)
\end{minipage} & \begin{minipage}[b]{\linewidth}\raggedright
$\Delta$ZPE
\end{minipage} & \begin{minipage}[b]{\linewidth}\raggedright
$\Delta C_{v,q}$
\end{minipage} \\
\midrule\noalign{}
\endfirsthead
\multicolumn{5}{@{}l}{\itshape Table \thetable{} (continued)}\\
\toprule\noalign{}
\begin{minipage}[b]{\linewidth}\raggedright
\kw{mt\_M}/\kw{mt\_P}
\end{minipage} & \begin{minipage}[b]{\linewidth}\raggedright
$M+P$
\end{minipage} & \begin{minipage}[b]{\linewidth}\raggedright
VACF drift (RMS)
\end{minipage} & \begin{minipage}[b]{\linewidth}\raggedright
$\Delta$ZPE
\end{minipage} & \begin{minipage}[b]{\linewidth}\raggedright
$\Delta C_{v,q}$
\end{minipage} \\
\midrule\noalign{}
\endhead
\bottomrule\noalign{}
\endlastfoot
16/16 & 32 & 0.82\% & +8.3\% & $-$12.0\% \\
32/16 & 48 & 0.73\% & +8.5\% & $-$12.2\% \\
32/32 & 64 & 0.67\% & +10.5\% & $-$14.9\% \\
64/64 & 128 & 0.50\% & +10.4\% & $-$15.3\% \\
128/128 & 256 & 0.29\% & +8.4\% & $-$11.4\% \\
256/256 & 512 & 0.12\% & +2.6\% & $-$3.4\% \\
\end{longtable}

\subsection{Complete Mode $\times$ Correlator Sweep}\label{complete-mode-correlator-sweep}

To verify that \emph{every} partitioning option reproduces the reference
post-processor with \emph{either} correlator, we ran the full matrix of six public analysis settings (\kw{1PT}; \kw{2PT} with \kw{rigorous}, \kw{lin2003},
\kw{desjarlais}, and \kw{r2pt} refinements; and \kw{3PT}), each with the FFT and the multi-tau correlator, on both a monatomic and a molecular system, against
\pyxPT{}. On each system the FFT fixes were declared concurrently and shared
one velocity buffer (Section~\ref{shared-velocity-buffers}), so the whole
matrix came from a single pass over the trajectory. The monatomic test
replays the liquid-argon trajectory of
Section~\ref{liquid-argon}; the molecular test
is the SPC/E water solvent of Section~\ref{dilute-propane-in-spce-water-same-trajectory-multi-tau-fft-and-py-xpt} (the
897-molecule, ensemble-averaged group), with \pyxPT{} analysing the identical
trajectory. Tables~\ref{tab:8} and~\ref{tab:9} collect the results.

\begin{longtable}[]{@{}
  >{\raggedright\arraybackslash}p{(\columnwidth - 8\tabcolsep) * \real{0.2639}}
  >{\raggedright\arraybackslash}p{(\columnwidth - 8\tabcolsep) * \real{0.1528}}
  >{\raggedright\arraybackslash}p{(\columnwidth - 8\tabcolsep) * \real{0.2361}}
  >{\raggedright\arraybackslash}p{(\columnwidth - 8\tabcolsep) * \real{0.1250}}
  >{\raggedright\arraybackslash}p{(\columnwidth - 8\tabcolsep) * \real{0.2222}}@{}}
\caption{Liquid argon ($\rho$* = 0.85, T* = 1.1): reduced molar entropy $S^*$ ($k_B$/atom) for all six settings, FFT and multi-tau correlators, vs \pyxPT{}.}\label{tab:8}\\
\toprule\noalign{}
\begin{minipage}[b]{\linewidth}\raggedright
mode / refinement
\end{minipage} & \begin{minipage}[b]{\linewidth}\raggedright
FFT $S^*$
\end{minipage} & \begin{minipage}[b]{\linewidth}\raggedright
multi-tau $S^*$
\end{minipage} & \begin{minipage}[b]{\linewidth}\raggedright
py-xPT
\end{minipage} & \begin{minipage}[b]{\linewidth}\raggedright
FFT vs py-xPT
\end{minipage} \\
\midrule\noalign{}
\endfirsthead
\multicolumn{5}{@{}l}{\itshape Table \thetable{} (continued)}\\
\toprule\noalign{}
\begin{minipage}[b]{\linewidth}\raggedright
mode / refinement
\end{minipage} & \begin{minipage}[b]{\linewidth}\raggedright
FFT $S^*$
\end{minipage} & \begin{minipage}[b]{\linewidth}\raggedright
multi-tau $S^*$
\end{minipage} & \begin{minipage}[b]{\linewidth}\raggedright
py-xPT
\end{minipage} & \begin{minipage}[b]{\linewidth}\raggedright
FFT vs py-xPT
\end{minipage} \\
\midrule\noalign{}
\endhead
\bottomrule\noalign{}
\endlastfoot
1PT & 6.7462 & 6.7462 & 6.7475 & $-$0.02\% \\
2PT / rigorous & 7.0820 & 7.0815 & 7.073 & +0.13\% \\
2PT / lin2003 & 7.5614 & 7.5611 & 7.553 & +0.11\% \\
2PT / desjarlais & 7.5656 & 7.5602 & 7.556 & +0.13\% \\
2PT / r2pt & 7.3659 & 7.3659 & 7.357 & +0.12\% \\
3PT & 7.3653 & 7.3650 & 7.362 & +0.04\% \\
\end{longtable}

\begin{longtable}[]{@{}
  >{\raggedright\arraybackslash}p{(\columnwidth - 8\tabcolsep) * \real{0.2639}}
  >{\raggedright\arraybackslash}p{(\columnwidth - 8\tabcolsep) * \real{0.1528}}
  >{\raggedright\arraybackslash}p{(\columnwidth - 8\tabcolsep) * \real{0.2361}}
  >{\raggedright\arraybackslash}p{(\columnwidth - 8\tabcolsep) * \real{0.1250}}
  >{\raggedright\arraybackslash}p{(\columnwidth - 8\tabcolsep) * \real{0.2222}}@{}}
\caption{SPC/E water solvent (897 molecules): per-molecule entropy $S_q$ (J mol\textsuperscript{-1} K\textsuperscript{-1}) for all six settings, FFT and multi-tau, vs \pyxPT{} (same trajectory).}\label{tab:9}\\
\toprule\noalign{}
\begin{minipage}[b]{\linewidth}\raggedright
mode / refinement
\end{minipage} & \begin{minipage}[b]{\linewidth}\raggedright
FFT $S_q$
\end{minipage} & \begin{minipage}[b]{\linewidth}\raggedright
multi-tau $S_q$
\end{minipage} & \begin{minipage}[b]{\linewidth}\raggedright
py-xPT
\end{minipage} & \begin{minipage}[b]{\linewidth}\raggedright
FFT vs py-xPT
\end{minipage} \\
\midrule\noalign{}
\endfirsthead
\multicolumn{5}{@{}l}{\itshape Table \thetable{} (continued)}\\
\toprule\noalign{}
\begin{minipage}[b]{\linewidth}\raggedright
mode / refinement
\end{minipage} & \begin{minipage}[b]{\linewidth}\raggedright
FFT $S_q$
\end{minipage} & \begin{minipage}[b]{\linewidth}\raggedright
multi-tau $S_q$
\end{minipage} & \begin{minipage}[b]{\linewidth}\raggedright
py-xPT
\end{minipage} & \begin{minipage}[b]{\linewidth}\raggedright
FFT vs py-xPT
\end{minipage} \\
\midrule\noalign{}
\endhead
\bottomrule\noalign{}
\endlastfoot
1PT & 52.877 & 52.886 & 52.736 & +0.27\% \\
2PT / rigorous & 56.859 & 56.825 & 56.843 & +0.03\% \\
2PT / lin2003 & 60.683 & 60.627 & 60.670 & +0.02\% \\
2PT / desjarlais & 60.805 & 60.727 & 60.791 & +0.02\% \\
2PT / r2pt & 57.925 & 57.852 & 57.915 & +0.02\% \\
3PT & 64.619 & 64.551 & 64.627 & $-$0.01\% \\
\end{longtable}

Across all twelve mode/correlator combinations on each system, \FixXPT{} with
the FFT correlator agrees with \pyxPT{} to within 0.15\% (monatomic) and 0.3\%
(molecular solvent; $\leq$ 0.05\% for the 2PT/3PT refinements, the 1PT baseline being
the largest at +0.27\%). The multi-tau correlator reproduces the FFT result to
within 0.15\% on every setting for these well-sampled groups (its only material
departure, the single-molecule solute translational channel, is documented in
Section~\ref{dilute-propane-in-spce-water-same-trajectory-multi-tau-fft-and-py-xpt}). The complete taxonomy, the harmonic
baseline, the four 2PT entropy refinements and the three-phase cage, is
therefore correct end-to-end and correlator-independent.

\subsection{Dilute Propane in SPC/E Water: Same-Trajectory Multi-tau, FFT, and py-xPT}\label{dilute-propane-in-spce-water-same-trajectory-multi-tau-fft-and-py-xpt}

The benzene benchmark (Section~\ref{liquid-benzene-polyatomic-mode-multi-tau-vs-fft-correlator}) validates the multi-tau correlator on
a neat liquid of identical rigid rotors. A more stringent molecular test
is a \emph{dilute solute} in a molecular solvent, where one group (the solute)
is sampled by a single molecule while the other (the solvent) is
ensemble-averaged over hundreds. A single propane molecule (11 atoms, so
all-atom OPLS~\cite{Jorgensen1996}) solvated in 897 SPC/E water molecules~\cite{Berendsen1987} (2702 atoms total,
\(T = 298\) K, \(V = 26\,903\) \AA{}\textsuperscript{3}, NVT, SHAKE-rigid water) was simulated for
20 ps of equilibration, after which one analysis window
(\(N_{\mathrm{every}} = 4\), \(N_{s} = 2048\), 8.2 ps) was processed three
ways on the identical frames: the \pyxPT{} post-processor (reading a
trajectory dumped every 4 steps), \FixXPT{} with \kw{correlator fft}, and
\FixXPT{} with \kw{correlator multitau} (\(mt\_M = 64\), \(mt\_P = 64\)). All
used \kw{mode} = 2PT (rigorous), C2v symmetry, and per-molecule
normalization. Table~\ref{tab:7} reports the per-molecule entropies of each group.
\begin{longtable}[]{@{}
  >{\raggedright\arraybackslash}p{(\columnwidth - 8\tabcolsep) * \real{0.2740}}
  >{\raggedright\arraybackslash}p{(\columnwidth - 8\tabcolsep) * \real{0.1370}}
  >{\raggedright\arraybackslash}p{(\columnwidth - 8\tabcolsep) * \real{0.1233}}
  >{\raggedright\arraybackslash}p{(\columnwidth - 8\tabcolsep) * \real{0.1918}}
  >{\raggedright\arraybackslash}p{(\columnwidth - 8\tabcolsep) * \real{0.2740}}@{}}
\caption{Per-molecule entropy \(S_{q}\) (J/mol/K) of the solute (propane) and solvent (water) groups, computed three ways on the same 8.2 ps window. Channel decomposition (trans/rot/vib) shown for each correlator.}\label{tab:7}\\
\toprule\noalign{}
\begin{minipage}[b]{\linewidth}\raggedright
Group
\end{minipage} & \begin{minipage}[b]{\linewidth}\raggedright
Channel
\end{minipage} & \begin{minipage}[b]{\linewidth}\raggedright
py-xPT
\end{minipage} & \begin{minipage}[b]{\linewidth}\raggedright
\FixXPT{} FFT
\end{minipage} & \begin{minipage}[b]{\linewidth}\raggedright
\FixXPT{} multi-tau
\end{minipage} \\
\midrule\noalign{}
\endfirsthead
\multicolumn{5}{@{}l}{\itshape Table \thetable{} (continued)}\\
\toprule\noalign{}
\begin{minipage}[b]{\linewidth}\raggedright
Group
\end{minipage} & \begin{minipage}[b]{\linewidth}\raggedright
Channel
\end{minipage} & \begin{minipage}[b]{\linewidth}\raggedright
py-xPT
\end{minipage} & \begin{minipage}[b]{\linewidth}\raggedright
\FixXPT{} FFT
\end{minipage} & \begin{minipage}[b]{\linewidth}\raggedright
\FixXPT{} multi-tau
\end{minipage} \\
\midrule\noalign{}
\endhead
\bottomrule\noalign{}
\endlastfoot
solute (propane) & \(S_{q}\) & 156.63 & 156.49 & 147.76 \\
& trans & --- & 64.07 & 55.78 \\
& rot & --- & 63.94 & 63.47 \\
& vib & --- & 28.47 & 28.51 \\
solvent (water) & \(S_{q}\) & 56.84 & 56.86 & 56.83 \\
& trans & --- & 46.42 & 46.39 \\
& rot & --- & 10.44 & 10.43 \\
\end{longtable}

For both groups, \FixXPT{} with the FFT correlator reproduces the
\pyxPT{} post-processor on the same trajectory to within 0.1\%
(solute \(-0.09\%\), solvent \(+0.04\%\)), confirming the same-trajectory
ground-truth equivalence of the on-the-fly molecular pipeline for a
mixture. The multi-tau correlator matches the FFT path to \(-0.06\%\) on
the well-averaged 897-molecule solvent, but differs by \(-5.6\%\) on the
single-molecule solute. The discrepancy is localized in the
\emph{translational} channel (\(S_{\mathrm{trans}}\) 64.1 $\rightarrow$ 55.8; the rotational
and vibrational channels agree to better than 1\%). The logarithmically
spaced lag grid of the telescoping correlator under-resolves the
long-time tail of the translational VACF when that tail is estimated from
a single molecule rather than an ensemble. This is a sampling effect that
vanishes under ensemble averaging, as the solvent demonstrates, rather
than a bias in the correlator, and it is the basis of the group-size
condition on the correlator choice (Section~\ref{on-line-multi-tau-correlator}).

\subsection{Supercooled TIP4P/2005 Water: Precision across independent runs}\label{supercooled-tip4p-2005-water}

The validations above all compare \FixXPT{} against \pyxPT{} on
identical velocity frames. As a fully independent check based on an independent
codebase, independent trajectory, and reference data prepared
independently of this implementation, we compare against the
\emph{original} stand-alone 2PT C++ post-processor~\cite{Ramji20232pt} on supercooled TIP4P/2005 water~\cite{Abascal2005}: 216
molecules (648 atoms) at \(T = 220\) K under NVT, deep in TIP4P/2005's
anomalous slow-dynamics regime, precisely the long-relaxation regime
that motivates the multi-tau correlator. The reference
values are the original-2PT analysis of a 10 ns block drawn from a
100 ns production trajectory.

\begin{figure}[htbp]
\centering
\includegraphics[width=\columnwidth]{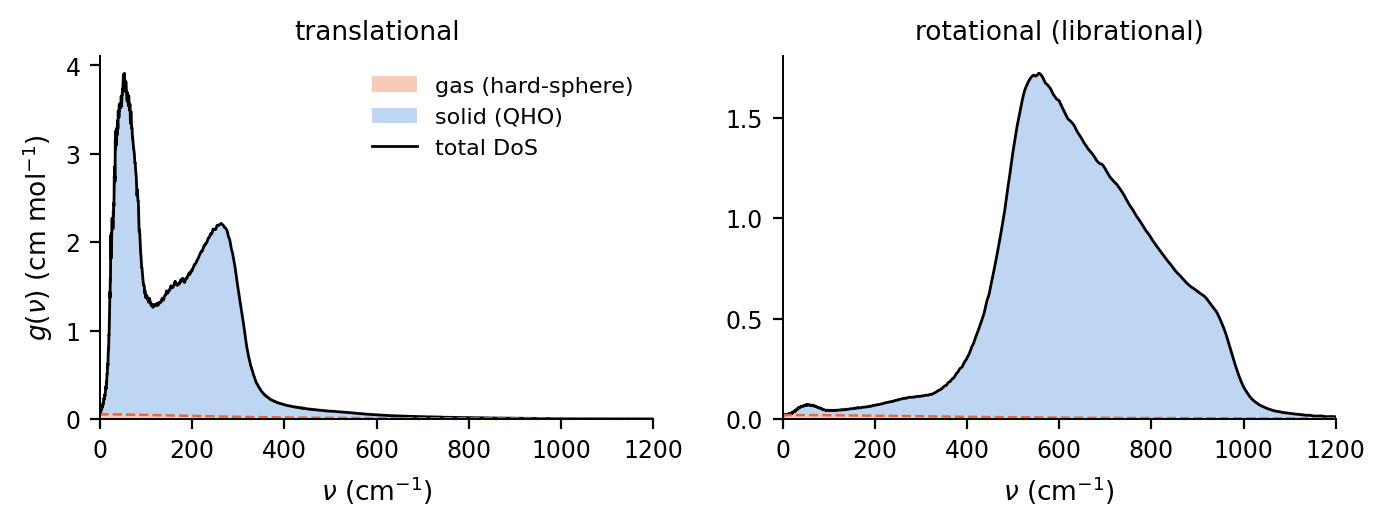}
\caption{Two-phase decomposition of the density of states of supercooled
TIP4P/2005 water at 220~K (averaged over 100~ps analysis windows of the
Table~\ref{tab:10} dataset). Left: translational channel, showing the
$\sim$50~cm$^{-1}$ O$\cdots$O$\cdots$O bending and $\sim$250~cm$^{-1}$
O$\cdots$O stretching bands; right: rotational (librational) channel.
The hard-sphere gas component (dashed) is barely visible on this scale:
at 220~K the fluidicity is only $f \approx 0.035$, i.e.\ the supercooled
liquid is almost entirely solid-like in the 2PT partition.}
\label{fig:dos}
\end{figure}

Our runs equilibrate independently from the same starting structure
(minimization, 10 ps heat-up) and then analyze one hundred consecutive
100 ps windows over 10 ns of production
(\(N_{\text{every}} = 2\), \(N_{s} = 50\,000\), molecular mode, C2v,
\kw{mode} = 2PT, \kw{correlator multitau} with
\kw{mt\_M} = \kw{mt\_P} = 64), repeated for three independently seeded replicas, 300 analysis windows in total. At this window length
the FFT correlator's velocity buffer would require \(\approx\)2.33 GB per
rank; the multi-tau replicas peaked below 270 MB per rank, a working
example of the correlator's intended use case (Section~\ref{memory-usage}). The
resulting density of states and its two-phase partition are shown in
Figure~\ref{fig:dos}.

\begin{longtable}[]{@{}
  >{\raggedright\arraybackslash}p{(\columnwidth - 6\tabcolsep) * \real{0.3000}}
  >{\raggedright\arraybackslash}p{(\columnwidth - 6\tabcolsep) * \real{0.2200}}
  >{\raggedright\arraybackslash}p{(\columnwidth - 6\tabcolsep) * \real{0.3000}}
  >{\raggedright\arraybackslash}p{(\columnwidth - 6\tabcolsep) * \real{0.1800}}@{}}
\caption{\FixXPT{} vs.\ the original 2PT post-processor for supercooled TIP4P/2005 water at 220 K, per molecule. Reference: original-2PT, 10 ns block of a 100 ns trajectory. \FixXPT{}: mean \(\pm\) s.d.\ over 300 consecutive 100 ps windows (three independently seeded and equilibrated 10 ns trajectories).}\label{tab:10}\\
\toprule\noalign{}
\begin{minipage}[b]{\linewidth}\raggedright
Quantity
\end{minipage} & \begin{minipage}[b]{\linewidth}\raggedright
original 2PT
\end{minipage} & \begin{minipage}[b]{\linewidth}\raggedright
\FixXPT{} (mean \(\pm\) \(\sigma\))
\end{minipage} & \begin{minipage}[b]{\linewidth}\raggedright
Deviation
\end{minipage} \\
\midrule\noalign{}
\endfirsthead
\multicolumn{4}{@{}l}{\itshape Table \thetable{} (continued)}\\
\toprule\noalign{}
\begin{minipage}[b]{\linewidth}\raggedright
Quantity
\end{minipage} & \begin{minipage}[b]{\linewidth}\raggedright
original 2PT
\end{minipage} & \begin{minipage}[b]{\linewidth}\raggedright
\FixXPT{} (mean \(\pm\) \(\sigma\))
\end{minipage} & \begin{minipage}[b]{\linewidth}\raggedright
Deviation
\end{minipage} \\
\midrule\noalign{}
\endhead
\bottomrule\noalign{}
\endlastfoot
$E_q$ {[}kJ/mol{]} & \(-40.14\) & \(-40.15 \pm 0.13\) & \(-0.03\%\) \\
$A_q$ {[}kJ/mol{]} & \(-47.81\) & \(-47.64 \pm 0.10\) & \(+0.4\%\) \\
$S_q$ {[}J/mol/K{]} & 34.91 & \(34.16 \pm 0.31\) & \(-2.2\%\) \\
$D$ {[}cm\textsuperscript{2}/s{]} & \(5.96\times10^{-7}\) & \((6.6 \pm 2.5)\times10^{-7}\) & within 1 s.d. \\
$f$ (fluidicity) & 0.0338 & \(0.0351 \pm 0.0081\) & within 1 s.d. \\
\end{longtable}

\begin{figure}[htbp]
\centering
\includegraphics[width=\columnwidth]{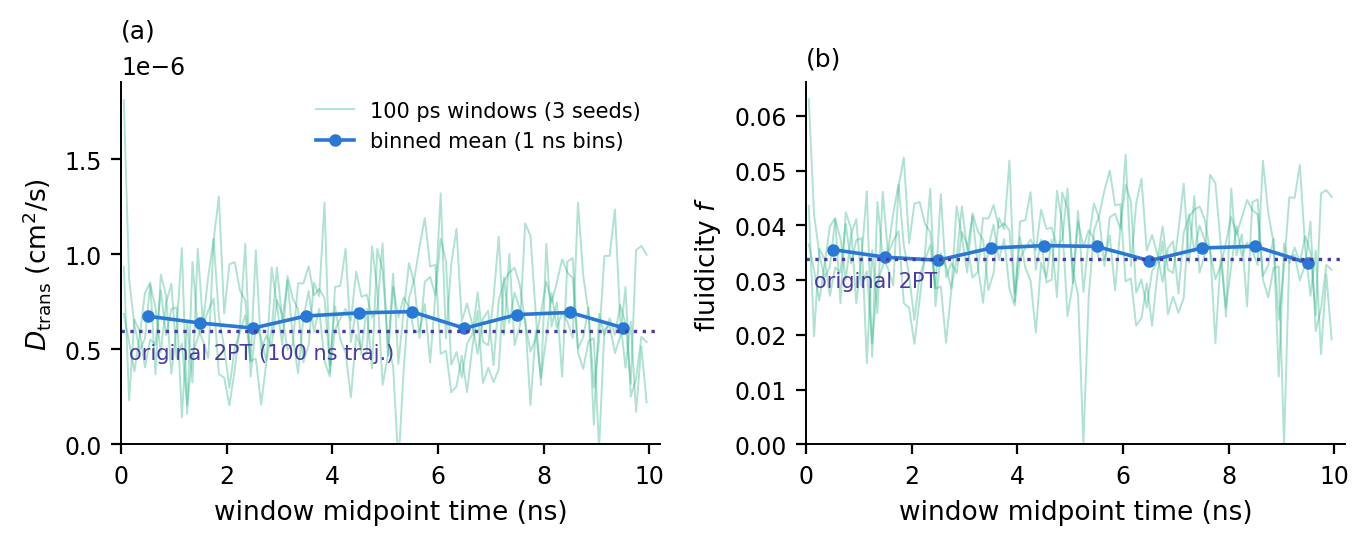}
\caption{Window-resolved (a) translational self-diffusivity and
(b) fluidicity of supercooled TIP4P/2005 water at 220~K. Thin traces:
one hundred consecutive 100~ps analysis windows for each of three
independently seeded 10~ns trajectories; circles: 1~ns binned means;
dotted line: the original-2PT reference. The window values are
stationary and bracket the reference from the first window onward.}
\label{fig:windows}
\end{figure}

The energies agree to reference precision ($E_q$ to
0.03\%, $A_q$ to 0.4\%) and the entropy to 2.2\%.
The transport-derived quantities $D$ and $f$, the noisiest 2PT outputs, scatter around the reference values, which lie within
one standard deviation of the window distribution for both.
Figure~\ref{fig:windows} resolves the individual windows in time: every
quantity is statistically stationary from the first 100 ps window
onward, across all three replicas and the full 10 ns, including the small $S_q$ offset, whose stability over 30 ns of
aggregate trajectory indicates the minor methodological difference
(window length, or convention details of the reference analysis) rather
than incomplete equilibration. The same analysis also quantifies a
practical hazard specific to this slow-relaxing regime: windows an order
of magnitude shorter (10 ps) overestimate both $D$ and $f$ by
2--7\(\times\), an error dominated by the short window's inability to
resolve the slowly decaying velocity correlations rather than by the age
of the underlying trajectory. Therefore, adequate window length is the operative requirement for \FixXPT{} of supercooled
liquids.

\subsection{Advanced Systems and Potentials}\label{multi-species-electrolytes-mlip}

Production use of \FixXPT{} frequently involves several concurrent
species-resolved fixes and, increasingly, machine-learned interatomic
potentials. We thus considered two advanced material classes:
\begin{itemize}
\item
\textbf{aqueous LiOH/KOH electrolyte}: 348 atoms; species groups
H\textsubscript{2}O, OH\textsuperscript{$-$}, Li\textsuperscript{+},
K\textsuperscript{+}; described entirely by classical non-polarizable
models: rigid TIP4P/2005 water~\cite{Abascal2005}, the Madrid-2019
scaled-charge parameters for Li\textsuperscript{+} and
K\textsuperscript{+}~\cite{Zeron2019} and the hydroxide model of
de~Lucas \emph{et al.}~\cite{deLucas2024}, both of which are
parameterized against TIP4P/2005 water.
\item 
\textbf{LiCoO\textsubscript{2} electrode}: 576 atoms; atomic Li,
Co, O groups; described by a Deep Potential neural-network
model~\cite{Zhang2018} as implemented in
\code{DeePMD-kit}~\cite{Wang2018}. The model uses the smooth-edition
two-body descriptor (\kw{se\_e2\_a}) and was trained on 6122 frames of
PBE-D2 data for
ground spin states of Li\textsubscript{x}CoO\textsubscript{2} across a
range of lithiation~\cite{Hu2023}. Each
system was run with three independent velocity seeds
(\(N_{\text{every}} = 2\), \(N_{s} = 2048\)), with four (electrolyte) or
three (electrode) species-resolved fixes active concurrently, and the
on-the-fly results compared against the C++~\cite{Ramji20232pt} and
\pyxPT{}~\cite{Ramji2026pyxpt} post-processors applied to the
identical dumped trajectory. 
\end{itemize}

Table~\ref{tab:11} summarizes the per-species
entropies: \FixXPT{} and the post-processed values agree to 0.04--2.7\%
(within \(\approx\)1.8 combined standard deviations in every case). The
largest deviation on both measures is electrode oxygen, at 2.7\% and
\(1.8\sigma\), where both estimates are individually precise. The electrode
rows constitute a direct demonstration of \FixXPT{} on a
machine-learned potential, where eliminating the velocity-trajectory
dump is particularly valuable: the force evaluation, not analysis I/O,
dominates a GPU-resident MLIP run (the GPU cost of the fix itself is
quantified in Section~\ref{gpu-scaling}).

\begin{longtable}[]{@{}
  >{\raggedright\arraybackslash}p{(\columnwidth - 8\tabcolsep) * \real{0.3400}}
  >{\raggedright\arraybackslash}p{(\columnwidth - 8\tabcolsep) * \real{0.2200}}
  >{\raggedright\arraybackslash}p{(\columnwidth - 8\tabcolsep) * \real{0.2200}}
  >{\raggedright\arraybackslash}p{(\columnwidth - 8\tabcolsep) * \real{0.2200}}@{}}
\caption{Per-species quantum entropy \(S_{q}\) (J/mol/K per molecule or atom, mean \(\pm\) s.d.\ over 3 velocity seeds): \FixXPT{} on-the-fly vs.\ post-processing on identical frames (the post-processed column is \pyxPT{}; the C++ post-processor~\cite{Ramji20232pt} gives identical values for the nonlinear and monatomic species).}\label{tab:11}\\
\toprule\noalign{}
\begin{minipage}[b]{\linewidth}\raggedright
Species (system)
\end{minipage} & \begin{minipage}[b]{\linewidth}\raggedright
\FixXPT{} FFT
\end{minipage} & \begin{minipage}[b]{\linewidth}\raggedright
\FixXPT{} multi-tau
\end{minipage} & \begin{minipage}[b]{\linewidth}\raggedright
post-processed
\end{minipage} \\
\midrule\noalign{}
\endfirsthead
\multicolumn{4}{@{}l}{\itshape Table \thetable{} (continued)}\\
\toprule\noalign{}
\begin{minipage}[b]{\linewidth}\raggedright
Species (system)
\end{minipage} & \begin{minipage}[b]{\linewidth}\raggedright
\FixXPT{} FFT
\end{minipage} & \begin{minipage}[b]{\linewidth}\raggedright
\FixXPT{} multi-tau
\end{minipage} & \begin{minipage}[b]{\linewidth}\raggedright
post-processed
\end{minipage} \\
\midrule\noalign{}
\endhead
\bottomrule\noalign{}
\endlastfoot
H\textsubscript{2}O (electrolyte) & \(75.44 \pm 0.17\) & \(75.36 \pm 0.17\) & \(75.41 \pm 0.17\) \\
OH\textsuperscript{$-$} (electrolyte) & \(62.63 \pm 0.55\) & \(62.50 \pm 0.53\) & \(62.22 \pm 0.51\) \\
K\textsuperscript{+} (electrolyte) & \(64.52 \pm 0.77\) & \(64.25 \pm 0.74\) & \(64.16 \pm 0.84\) \\
Li\textsuperscript{+} (electrolyte) & \(22.75 \pm 0.65\) & \(23.10 \pm 0.59\) & \(22.20 \pm 1.13\) \\
Li (electrode, DeepMD) & \(20.23 \pm 0.19\) & \(19.80 \pm 0.18\) & \(19.89 \pm 0.19\) \\
Co (electrode, DeepMD) & \(23.95 \pm 0.42\) & \(23.80 \pm 0.42\) & \(23.96 \pm 0.37\) \\
O (electrode, DeepMD) & \(15.03 \pm 0.16\) & \(14.58 \pm 0.14\) & \(14.63 \pm 0.16\) \\
\end{longtable}

\section{Performance}\label{performance}

\subsection{Computational Overhead}\label{computational-overhead}

The dominant cost of \FixXPT{} is the FFT pipeline in \code{run\_analysis()},
executed once per analysis window; between windows the fix adds only the
per-step velocity-frame push and the energy/pressure tally.

To quantify the end-to-end cost we ran two canonical systems, liquid argon (500 atoms, monatomic, \kw{real} units) and SPC/E water (molecular mode), with the fix disabled (\kw{off}) and enabled (\kw{on}), on 1 and 4
MPI ranks of a single workstation, averaging over replicate runs.
Table~\ref{tab:12} reports the mean LAMMPS loop time, the relative wall-clock
overhead of enabling the fix, the fraction of loop time spent inside the
fix (the LAMMPS \code{Modify} class), and the mean peak resident-set size.

\begin{longtable}[]{@{}
  >{\raggedright\arraybackslash}p{(\columnwidth - 14\tabcolsep) * \real{0.0842}}
  >{\raggedright\arraybackslash}p{(\columnwidth - 14\tabcolsep) * \real{0.0421}}
  >{\raggedright\arraybackslash}p{(\columnwidth - 14\tabcolsep) * \real{0.1684}}
  >{\raggedright\arraybackslash}p{(\columnwidth - 14\tabcolsep) * \real{0.1579}}
  >{\raggedright\arraybackslash}p{(\columnwidth - 14\tabcolsep) * \real{0.1053}}
  >{\raggedright\arraybackslash}p{(\columnwidth - 14\tabcolsep) * \real{0.1158}}
  >{\raggedright\arraybackslash}p{(\columnwidth - 14\tabcolsep) * \real{0.1684}}
  >{\raggedright\arraybackslash}p{(\columnwidth - 14\tabcolsep) * \real{0.1579}}@{}}
\caption{\FixXPT{} run-time and memory overhead (default FFT correlator, replicated layout) relative to plain MD, for liquid argon (25\,600 steps) and SPC/E water (60\,000 steps) on 1 and 4 MPI ranks of a single workstation. Each cell is the mean over six replicates, with the fix-on and fix-off runs of each replicate executed back-to-back (interleaved) so that slow drift in background load cancels out of the on/off difference; per-run s.d. \textless{} 2\% except water at \code{np} = 4 (\(\approx\)8\%). Peak RSS is measured by wrapping each run in GNU \code{/usr/bin/time -v}: under MPI this reports the single largest reaped rank's peak resident set (\code{getrusage(RUSAGE\_CHILDREN)}), not a sum across ranks. This is the appropriate per-rank quantity for buffer sizing; in this layout node totals are \code{np} times larger (Section~\ref{mpi-scaling}).}\label{tab:12}\\
\toprule\noalign{}
\begin{minipage}[b]{\linewidth}\raggedright
System
\end{minipage} & \begin{minipage}[b]{\linewidth}\raggedright
np
\end{minipage} & \begin{minipage}[b]{\linewidth}\raggedright
Loop \kw{off} (s)
\end{minipage} & \begin{minipage}[b]{\linewidth}\raggedright
Loop \kw{on} (s)
\end{minipage} & \begin{minipage}[b]{\linewidth}\raggedright
Overhead
\end{minipage} & \begin{minipage}[b]{\linewidth}\raggedright
Fix share
\end{minipage} & \begin{minipage}[b]{\linewidth}\raggedright
RSS \kw{off} (MB)
\end{minipage} & \begin{minipage}[b]{\linewidth}\raggedright
RSS \kw{on} (MB)
\end{minipage} \\
\midrule\noalign{}
\endfirsthead
\multicolumn{8}{@{}l}{\itshape Table \thetable{} (continued)}\\
\toprule\noalign{}
\begin{minipage}[b]{\linewidth}\raggedright
System
\end{minipage} & \begin{minipage}[b]{\linewidth}\raggedright
np
\end{minipage} & \begin{minipage}[b]{\linewidth}\raggedright
Loop \kw{off} (s)
\end{minipage} & \begin{minipage}[b]{\linewidth}\raggedright
Loop \kw{on} (s)
\end{minipage} & \begin{minipage}[b]{\linewidth}\raggedright
Overhead
\end{minipage} & \begin{minipage}[b]{\linewidth}\raggedright
Fix share
\end{minipage} & \begin{minipage}[b]{\linewidth}\raggedright
RSS \kw{off} (MB)
\end{minipage} & \begin{minipage}[b]{\linewidth}\raggedright
RSS \kw{on} (MB)
\end{minipage} \\
\midrule\noalign{}
\endhead
\bottomrule\noalign{}
\endlastfoot
argon & 1 & 4.64 & 5.61 & +21\% & 9.0\% & 31.9 & 42.3 \\
argon & 4 & 1.71 & 2.17 & +27\% & 14.1\% & 32.2 & 42.3 \\
water & 1 & 289.5 & 341.7 & +18\% & 5.6\% & 42.0 & 339.8 \\
water & 4 & 91.7 & 113.3 & +24\% & 13.8\% & 40.4 & 338.3 \\
\end{longtable}

The fix adds 18--21\% wall time at \code{np} = 1, rising to 24--27\% at
\code{np} = 4. In the replicated layout of these runs the overhead's
\emph{relative} share grows with rank count, because every rank receives
and stores each complete frame, a cost that does not shrink as the
force and neighbor work does. The distributed layout divides that cost
across the ranks, and its overhead does not grow with rank count
(Section~\ref{mpi-scaling}). In both
systems the fix's own share of the loop (the \code{Modify} column) is
5.6--14.1\%; the
remainder of the overhead is the per-step energy/pressure tallying the
fix schedules via \code{addstep} (Section~\ref{compute-scheduling-the-addstep-pattern}). The per-window analysis itself
is dominated by the FFT pipeline in \code{run\_analysis()}, which
transforms each unit in each Cartesian direction: for the benzene window
that is $3(64+64+768) = 2688$ forward transforms across the
translational, rotational and vibrational channels, followed by three
inverse transforms and three cosine transforms. The Desjarlais $B_g$ refinement adds less than 1 ms
per channel per window. 

The workflow \FixXPT{} replaces is the more informative baseline. The
liquid benzene system of Section~\ref{liquid-benzene-polyatomic-mode-multi-tau-vs-fft-correlator} was re-run from one restart
file (4 MPI ranks, $N_{every}$ = 5, 100 ps NVT) as a baseline NVT dumping
velocities every fifth step, and with each correlator in turn
(Table~\ref{tab:13}). The FFT path adds $\sim$8\% wall over that dump
workflow and the multi-tau path $\sim$30\%, trading a per-step push for
the once-per-window FFT. Either eliminates the 1.33 GB trajectory,
reducing on-disk output $\sim$320$\times$. At shorter windows
the two correlators converge in cost: on SPC/E water (\(N_{s} = 5000\),
\code{np} = 1, six replicates on a dedicated node) they add
\(11.4 \pm 0.4\%\) and \(12.5 \pm 0.4\%\) over plain MD, while the fix's
peak-memory increment differs eighteen-fold (\(+298\) MB vs \(+17\) MB,
from the water row of Table~\ref{tab:15}). The choice between correlators is
therefore governed by memory, not speed, except at the longest windows.

\begin{longtable}[]{@{}
  >{\raggedright\arraybackslash}p{(\columnwidth - 4\tabcolsep) * \real{0.5909}}
  >{\raggedright\arraybackslash}p{(\columnwidth - 4\tabcolsep) * \real{0.1364}}
  >{\raggedright\arraybackslash}p{(\columnwidth - 4\tabcolsep) * \real{0.2500}}@{}}
\caption{Total LAMMPS wall time for one 100 ps NVT production run of liquid benzene at 298 K (64 molecules, 4 MPI ranks, $N_{every}$ = 5, $N_s$ = 20 000, replicated layout). All runs load the same restart file.}\label{tab:13}\\
\toprule\noalign{}
\begin{minipage}[b]{\linewidth}\raggedright
Run mode
\end{minipage} & \begin{minipage}[b]{\linewidth}\raggedright
Wall time
\end{minipage} & \begin{minipage}[b]{\linewidth}\raggedright
On-disk output
\end{minipage} \\
\midrule\noalign{}
\endfirsthead
\multicolumn{3}{@{}l}{\itshape Table \thetable{} (continued)}\\
\toprule\noalign{}
\begin{minipage}[b]{\linewidth}\raggedright
Run mode
\end{minipage} & \begin{minipage}[b]{\linewidth}\raggedright
Wall time
\end{minipage} & \begin{minipage}[b]{\linewidth}\raggedright
On-disk output
\end{minipage} \\
\midrule\noalign{}
\endhead
\bottomrule\noalign{}
\endlastfoot
Baseline NVT + trajectory dump (no \FixXPT{}) & 0:50 & 1.33 GB lammpstrj \\
NVT + \kw{fix xpt \ldots\ correlator fft} & 0:54 & 4.24 MB triplet \\
NVT + \kw{fix xpt \ldots\ correlator multitau} & 1:05 & 4.27 MB triplet \\
\end{longtable}

The same comparison was repeated across system sizes and both material
classes of Section~\ref{multi-species-electrolytes-mlip}: full 75\,000-step trajectories, one
workstation, the C++ post-processor~\cite{Ramji20232pt} as the
external route (Table~\ref{tab:14}). Wherever the on-the-fly fix fits in memory it
is faster end-to-end than dump-plus-post-processing, by
1.26--1.45\(\times\). The external route must write the
trajectory (dominant for the DeepMD electrode, where the dump run alone
exceeded the entire on-the-fly run) and then read and re-transform it
(growing from 4.5 s to 1755 s across the size range; the electrode-96
row is a statistical tie, its \(\sim\)1000 s runs being MD-dominated and
subject to GPU run-to-run variation). This comparison flips at scale when
every rank holds the whole history: at 22\,272 atoms with four
species-resolved FFT-correlator fixes, the replicated per-rank velocity
buffers exhaust a 78 GB workstation at every practical rank count, and the
dump-plus-post-process route becomes the only viable one. The multi-tau
correlator (Section~\ref{memory-usage}) shrinks each history, and the
distributed layout (Section~\ref{mpi-scaling}) stores one copy of it across
the ranks rather than one per rank.

\begin{longtable}[]{@{}
  >{\raggedright\arraybackslash}p{(\columnwidth - 8\tabcolsep) * \real{0.3200}}
  >{\raggedright\arraybackslash}p{(\columnwidth - 8\tabcolsep) * \real{0.2000}}
  >{\raggedright\arraybackslash}p{(\columnwidth - 8\tabcolsep) * \real{0.2600}}
  >{\raggedright\arraybackslash}p{(\columnwidth - 8\tabcolsep) * \real{0.2200}}@{}}
\caption{End-to-end wall time: NVT with \FixXPT{} (on-the-fly, replicated layout) vs.\ NVT + velocity dump + external post-processing, full 75\,000-step trajectories on one workstation.}\label{tab:14}\\
\toprule\noalign{}
\begin{minipage}[b]{\linewidth}\raggedright
System
\end{minipage} & \begin{minipage}[b]{\linewidth}\raggedright
on-the-fly (s)
\end{minipage} & \begin{minipage}[b]{\linewidth}\raggedright
dump + post-proc.\ (s)
\end{minipage} & \begin{minipage}[b]{\linewidth}\raggedright
external / inline
\end{minipage} \\
\midrule\noalign{}
\endfirsthead
\multicolumn{4}{@{}l}{\itshape Table \thetable{} (continued)}\\
\toprule\noalign{}
\begin{minipage}[b]{\linewidth}\raggedright
System
\end{minipage} & \begin{minipage}[b]{\linewidth}\raggedright
on-the-fly (s)
\end{minipage} & \begin{minipage}[b]{\linewidth}\raggedright
dump + post-proc.\ (s)
\end{minipage} & \begin{minipage}[b]{\linewidth}\raggedright
external / inline
\end{minipage} \\
\midrule\noalign{}
\endhead
\bottomrule\noalign{}
\endlastfoot
electrode-96 (DeepMD) & 1203 & 1046 & 0.87 \\
electrolyte-348 & 130 & 164 & 1.26 \\
electrolyte-2784 & 441 & 571 & 1.29 \\
electrode-576 (DeepMD) & 3082 & 4456 & 1.45 \\
electrolyte-22\,272 & OOM & 4214 & --- \\
\end{longtable}

\subsection{Memory Usage}\label{memory-usage}

The velocity buffer requires $N_s \times \lceil N/P \rceil \times 3 \times 8$ bytes of RAM
per rank in the distributed layout ($N_s \times N_{\mathrm{max\_tag}} \times 3 \times 8$ bytes
replicated). For 1000 atoms and $N_s$ = 512 on one rank, this is
approximately 12 MB. In molecular mode, four additional
buffers are allocated: three of size proportional to
$N_{mol} \times N_s$ (centre-of-mass velocity, scaled angular velocity and
lab-frame angular velocity) and the per-atom vibrational residual, which is
proportional to $N_{atoms} \times N_s$ and is the largest of the four. Molecular
mode also holds the map from atom tag to molecule index, one integer per tag
on every rank, rebuilt at each window start: 4 bytes $\times N_{\mathrm{max\_tag}}$,
166 kB at 41\,472 atoms, and the one structure whose size does not fall with
the rank count. The \code{memory\_usage()} callback
reports the per-rank contribution to the LAMMPS memory summary.

One practical caveat modifies this estimate substantially: the
zero-padded transform length \(2N_{s}\) determines which FFT algorithm
the backend selects, and a length that is not a power of two can select
a plan whose workspace inflates the correlator's footprint by roughly an
order of magnitude, measured at \(\approx\)9\(\times\) for
\(N_{s} = 18\,750\) (5.6 GB where the buffer formula predicts far less).
Powers of two avoid this entirely and are strongly
recommended for the FFT correlator.

The peak-RSS columns of Table~\ref{tab:12} show this cost directly. For the
500-atom argon run the FFT correlator adds only $\sim$10 MB (32 $\rightarrow$ 42 MB,
independent of rank count). For the SPC/E water run the buffer dominates:
enabling the fix raises peak RSS from $\sim$42 MB to $\sim$340 MB, eight times the baseline, because the molecular pipeline holds the full
translational, rotational, and vibrational velocity histories for the
window. This linear-in-$N_s$ growth is the memory bottleneck the
multi-tau correlator is designed to remove.

Because $N_s$ enters linearly, long windows make this buffer the
LAMMPS-level RSS: on the benzene benchmark ($N_s$ = 20 000, 768 atoms)
the FFT correlator peaked at $\sim$1.14 GB against $\sim$52 MB for the
baseline NVT run, while the multi-tau correlator, whose footprint grows
only logarithmically with $N_s$ (Section~\ref{on-line-multi-tau-correlator}), peaked at
$\sim$96 MB on the same window.
The benzene case varies the window length at fixed system size;
Table~\ref{tab:15} varies the \emph{system size} at fixed window configuration
(molecular SPC/E water, \(N_{s} = 5000\), single rank, peak RSS via
\code{/usr/bin/time -v}; monatomic argon shown for contrast). The FFT
correlator's footprint reaches the GB scale by \(\approx\)5000 atoms,
and the multi-tau reduction factor \emph{grows} with atom count
(5.8\(\times\) \(\rightarrow\) 13.5\(\times\)): the advantage compounds
exactly where absolute memory pressure is largest. For the monatomic argon case (small \(N_{s}\), no molecular channel buffers) the two correlators differ by only \(\approx\)10\%, consistent with the
window-length buffer being a minor contributor there.

\begin{longtable}[]{@{}
  >{\raggedright\arraybackslash}p{(\columnwidth - 8\tabcolsep) * \real{0.3200}}
  >{\raggedright\arraybackslash}p{(\columnwidth - 8\tabcolsep) * \real{0.1400}}
  >{\raggedright\arraybackslash}p{(\columnwidth - 8\tabcolsep) * \real{0.1800}}
  >{\raggedright\arraybackslash}p{(\columnwidth - 8\tabcolsep) * \real{0.2000}}
  >{\raggedright\arraybackslash}p{(\columnwidth - 8\tabcolsep) * \real{0.1600}}@{}}
\caption{Peak resident set size of the two correlators versus system size (\code{np} = 1, single analysis window). Water rows share \(N_{s} = 5000\) and molecular mode; the larger cell is the same SPC/E box replicated \(2\times2\times2\).}\label{tab:15}\\
\toprule\noalign{}
\begin{minipage}[b]{\linewidth}\raggedright
System
\end{minipage} & \begin{minipage}[b]{\linewidth}\raggedright
atoms
\end{minipage} & \begin{minipage}[b]{\linewidth}\raggedright
FFT (MB)
\end{minipage} & \begin{minipage}[b]{\linewidth}\raggedright
multi-tau (MB)
\end{minipage} & \begin{minipage}[b]{\linewidth}\raggedright
Reduction
\end{minipage} \\
\midrule\noalign{}
\endfirsthead
\multicolumn{5}{@{}l}{\itshape Table \thetable{} (continued)}\\
\toprule\noalign{}
\begin{minipage}[b]{\linewidth}\raggedright
System
\end{minipage} & \begin{minipage}[b]{\linewidth}\raggedright
atoms
\end{minipage} & \begin{minipage}[b]{\linewidth}\raggedright
FFT (MB)
\end{minipage} & \begin{minipage}[b]{\linewidth}\raggedright
multi-tau (MB)
\end{minipage} & \begin{minipage}[b]{\linewidth}\raggedright
Reduction
\end{minipage} \\
\midrule\noalign{}
\endhead
\bottomrule\noalign{}
\endlastfoot
argon (monatomic, \(N_{s}\) = 512) & 500 & 42.8 & 38.4 & 1.1$\times$ \\
SPC/E water & 648 & 340.5 & 58.5 & 5.8$\times$ \\
SPC/E water \(2\times2\times2\) & 5184 & 2438 & 181 & 13.5$\times$ \\
\end{longtable}

\subsection{MPI Scaling}\label{mpi-scaling}

In the distributed layout every per-frame cost of \FixXPT{} divides
across the ranks. A sample frame costs one \code{MPI\_Alltoall} of counts
and one \code{MPI\_Alltoallv} carrying $\mathcal{O}(N/P)$ datums per rank
(Section~\ref{tag-based-parallel-atom-migration}), each rank decomposes
and stores only its home atoms and molecules, and the window-end
transforms run over the same home units before one reduction assembles
the power spectra (Section~\ref{mpi-parallel-correctness}). The replicated
layout instead delivers each frame whole to every rank, which decomposes
and stores all of it, so the per-rank work of a frame stays
$\mathcal{O}(N)$ while the force work falls as $1/P$.

\begin{figure}[htbp]
\centering
\includegraphics[width=\columnwidth]{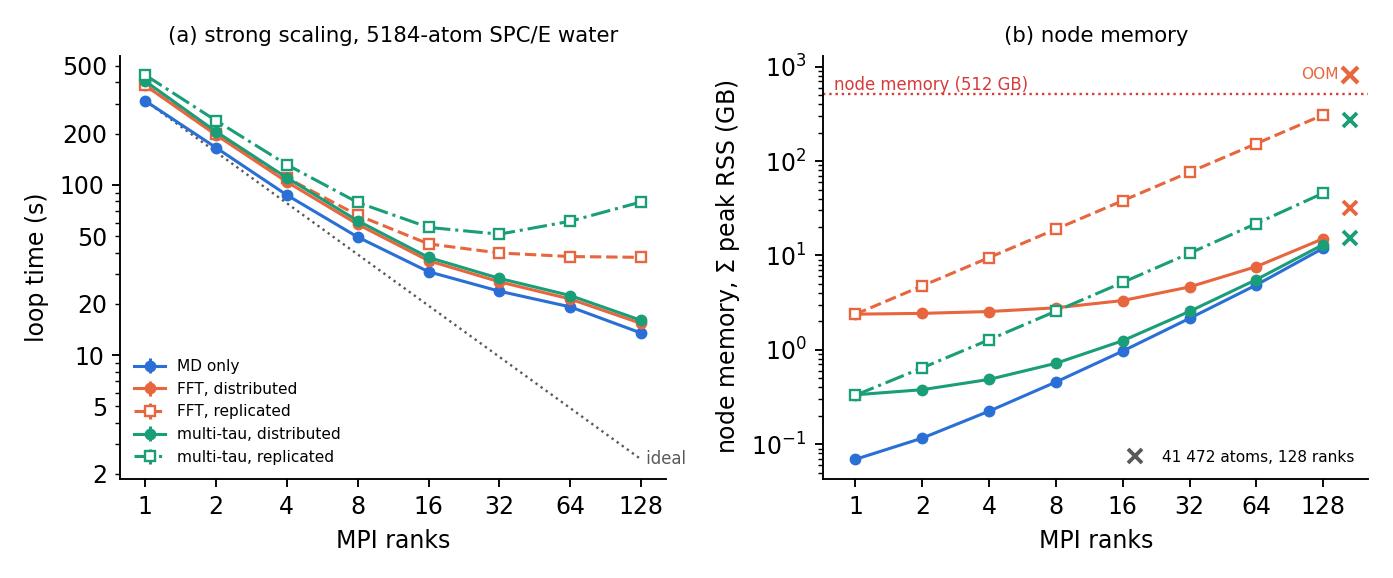}
\caption{Single-node scaling of \FixXPT{} on a 128-core Perlmutter CPU
node (5184-atom SPC/E water, molecular mode, one analysis window of
$N_s = 5000$; mean of two replicates, with bars spanning them).
(a)~Loop time versus MPI rank count for the MD alone and with the FFT or
multi-tau correlator in the distributed (filled) or replicated (open)
layout; the dotted line is ideal scaling of the MD from one rank.
(b)~Node memory, the sum over ranks of each rank's peak resident set.
The distributed layout stores one copy of the window's history across
the node and the replicated layout one copy per rank. Crosses mark
41\,472 atoms on 128 ranks, a single run, where the replicated FFT run
exceeds the node's 512~GB.}
\label{fig:scaling}
\end{figure}

We measured both layouts on one CPU node of the Perlmutter system at the
National Energy Research Scientific Computing Center, which carries two
AMD EPYC 7763 processors (128 physical cores) and 512 GB of memory. The
system is the 5184-atom SPC/E water of Table~\ref{tab:15} in molecular
mode, with one analysis window of $N_s = 5000$ frames sampled every
second step, run on 1--128 ranks at one rank per core. Each point is the
mean of two replicates run on separate nodes, which differ by 0.5\% at
the median, and node memory is the sum over ranks of each rank's peak
resident set, measured with GNU \code{time} (Figure~\ref{fig:scaling}).

In the distributed layout the fix follows the MD. The FFT-correlator run
adds 23\% to the loop time on one rank and 11--20\% on 2--128 ranks, and
it speeds up 25-fold from 1 to 128 ranks, against 23-fold for the MD
alone (15.3 against 13.5 s on 128 ranks). The multi-tau run stays within
6\% of it throughout. The replicated FFT run matches the distributed
one on one and two ranks, but both replicated runs fall behind as ranks
are added, because their per-rank work does not shrink. The FFT run
levels off near 38 s from 32 ranks on, 2.4 times the distributed loop at
128 ranks, and the multi-tau run is fastest on 32 ranks and slows to
79 s on 128.

Node memory separates the layouts more sharply
(Figure~\ref{fig:scaling}b). The distributed FFT run holds one copy of
the window's history across the node, so beyond that copy its footprint
grows only with that of the MD, reaching 15.1 GB on 128 ranks, of which
11.9 GB is the MD's own. The replicated FFT run holds a full copy on
every rank, 2.4 GB each, and reaches 309 GB on 128 ranks. At 41\,472
atoms (the 648-atom box of Table~\ref{tab:15} replicated
$4\times4\times4$) on 128 ranks, run once rather than twice, the
distributed FFT run completes in 31.8 GB, whereas the replicated run
exceeds the node's 512 GB and is killed. The multi-tau correlator shrinks each copy by an order of
magnitude (Table~\ref{tab:15}), so its replicated run still fits at this
size (273 GB), but at 8 times the loop time of the distributed run (571
against 68 s). Only the distributed layout removes the per-rank copy.

\subsection{GPU Scaling}\label{gpu-scaling}

\begin{figure}[htbp]
\centering
\includegraphics[width=\columnwidth]{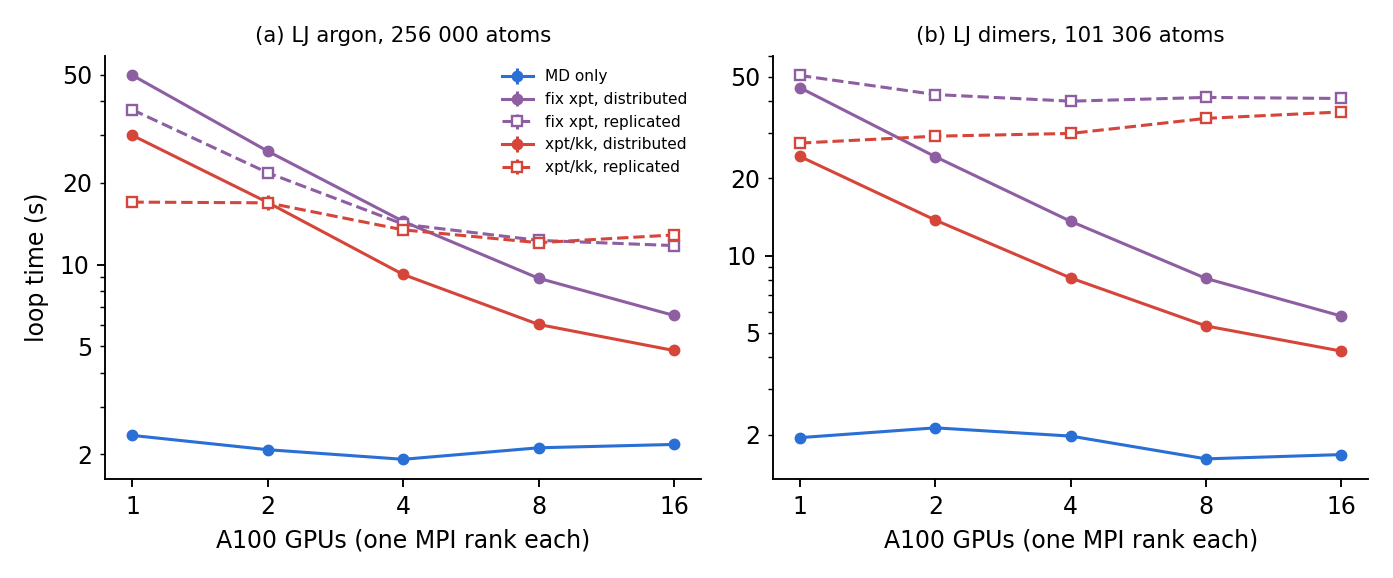}
\caption{GPU strong scaling of \FixXPT{} on Perlmutter A100 nodes (one MPI
rank per GPU, four GPUs per node, GPU-aware MPI, MD under \kw{-sf kk};
$N_s = 1000$; mean of two replicates, with bars spanning them). (a)~256\,000
Lennard-Jones argon atoms, monatomic mode. (b)~101\,306 atoms of
Lennard-Jones dimers, molecular mode. Loop time for the MD alone and with
the CPU class (\kw{fix xpt}) or \kw{xpt/kk} in the distributed (filled) or
replicated (open) layout.}
\label{fig:gpu-scaling}
\end{figure}

We measured the cost of the fix on the GPU nodes of Perlmutter, each
carrying one AMD EPYC 7763 processor and four NVIDIA A100 (40 GB) GPUs. The
MD ran under the KOKKOS package (\kw{-k on g 4 -sf kk}) with one MPI rank
per GPU and GPU-aware MPI, on 1--16 GPUs across 1--4 nodes. Two systems
cover the two analysis modes: 256\,000 Lennard-Jones argon atoms at 100~K
in monatomic mode (5000 steps, $N_s = 1000$ frames sampled every fifth
step) and 50\,653 Lennard-Jones dimers (101\,306 atoms) at 90~K in
molecular mode (4000 steps, $N_s = 1000$ frames sampled every fourth
step). Each system ran without the fix, with the CPU class (\kw{fix xpt},
fed from host copies of the atom arrays) and with \kw{xpt/kk}, in both
layouts, as two replicates that differ by 0.7\% at the median
(Figure~\ref{fig:gpu-scaling}).

At these sizes the MD alone does not strong-scale: its loop holds at
1.6--2.4 s on 1--16 GPUs. The figure therefore resolves the fix's own
cost, the loop time less that of the fix-off run. In the distributed
layout this cost falls 8.8--11-fold from one GPU to sixteen, for argon
from 27.6 to 2.7 s with \kw{xpt/kk} (47.4 to 4.3 s with the CPU class)
and for the dimers from 22.4 to 2.6 s (43.3 to 4.1 s). At every GPU
count \kw{xpt/kk} costs 0.52--0.62 of the CPU class. In the replicated
layout the cost stops falling. For argon it levels off at 9.6--12 s from
four GPUs on with either class. For the dimers it falls once, from 48.6 s
on one GPU to 38--40 s on two, and then holds there with the CPU class
while rising from 26 to 35 s with \kw{xpt/kk}, because every rank
assembles and decomposes all 50\,653 molecules at every frame.

Two limits bound these numbers. First, on one GPU, where both layouts
hold the full history, the replicated layout is the faster one for
argon (17.0 against 29.9 s with \kw{xpt/kk}, 37.2 against 49.8 s with the
CPU class). It indexes each frame by tag directly and, under \kw{xpt/kk},
keeps it on the device, whereas the distributed layout routes every atom
through the home-rank exchange and a binary search for its slot. For the
dimers the distributed layout is faster even on one GPU. On a single rank
the fix therefore selects the replicated layout for atomic groups and keeps
the distributed one in molecular mode, unless \kw{buffer\_layout} is set
explicitly. Second, against a force field that the GPU makes nearly free,
the fix remains the larger part of the loop even on sixteen GPUs (2.7 s
against 2.2 s for the argon MD). A fix cost of this size is small beside
the force evaluation of a machine-learned potential, whose runs are
dominated by the MD (Table~\ref{tab:14}).

\section{Conclusions}\label{conclusions}

In summary, we have presented \FixXPT{}, a native LAMMPS C++ implementation of the
Two-Phase Thermodynamics method that eliminates the need for
post-processing of velocity trajectory files. We resolved numerous
non-trivial conflicts between xPT's algorithmic requirements and
LAMMPS's runtime environment. The molecular decomposition mode, with
per-frame inertia-tensor diagonalization, extends 2PT to polyatomic
systems and is validated against published results and, on identical
velocity frames, against the stand-alone \pyxPT{} post-processor across
every analysis mode, correlator and system studied here. The in-line
multi-tau correlator extends the on-the-fly workflow to long analysis
windows where the FFT correlator's velocity buffer would dominate the
per-rank memory footprint, reducing peak memory by an order of magnitude
while reproducing the 2PT thermodynamics of the state variables. Both
correlators eliminate the gigabyte-scale on-disk
velocity trajectory required by the conventional post-processing
workflow. Distributing each window's history across the MPI ranks keeps
the memory the fix adds to a node nearly independent of the rank count and lets
its cost fall with the rank count on CPUs and GPUs alike.

\FixXPT{} is distributed as a drop-in XPT package that builds within any
recent LAMMPS source tree. It should apply directly to the thermodynamics
of arbitrary systems and potential energy functions, including
machine-learned interatomic potentials, polarizable force
fields and coarse-grained models. Future work will include adaptive
sampling strategies that adjust $N_s$ based on VACF convergence.

More generally, moving the density-of-states analysis inside the
integrator changes what a computed entropy/free energy can be used for. A quantity that costs a
separate post-processing stage is measured once, at the end. A quantity
that costs 10--30\% of the loop can be monitored continuously, resolved
by species or by sub-volume, and used to steer the simulation while it
runs, provided the window resolves the slowest velocity
correlations (Section~\ref{supercooled-tip4p-2005-water}). The thermodynamic quantities
that are hardest to obtain thereby become running observables of the
simulation itself, not yet as cheap as the temperature or the
pressure, but read out from the same run.

\section*{CRediT authorship contribution statement}
%% One unbreakable box: the statement can neither split across a page nor
%% strand the last author's roles, wherever the text above reflows.
\noindent\begin{minipage}{\linewidth}
\textbf{Robert Ramji:} Data curation, Investigation, Methodology, Software,
Validation, Visualization, Writing -- original draft.
\textbf{Ilya Mazalov:} Investigation, Software, Validation, Writing -- review
\& editing.
\textbf{Ricardo Buarque:} Formal analysis, Methodology, Writing -- review \&
editing.
\textbf{Alexandria Do:} Investigation, Validation, Writing -- review \& editing.
\textbf{Xiaoxu Ruan:} Investigation, Validation, Writing -- review \& editing.
\textbf{Tod A. Pascal:} Conceptualization, Funding acquisition, Project
administration, Resources, Supervision, Writing -- original draft, Writing --
review \& editing.
\end{minipage}

\section*{Acknowledgments}

We thank Shiang-Tai Lin and William A. Goddard III for the original C++
2PT post-processor. This work was supported by the National Science
Foundation under grant CBET-2145144. Computational resources were
provided by the San Diego Supercomputer Center (SDSC) Expanse cluster
through ACCESS allocation TG-CHE230014. This research also used resources
of the National Energy Research Scientific Computing Center (NERSC), a
U.S.\ Department of Energy Office of Science User Facility located at
Lawrence Berkeley National Laboratory, operated under Contract
No.~DE-AC02-05CH11231 using NERSC award DDR-ERCAP0039121.

\section*{Declaration of Competing Interest}

The authors declare that they have no known competing financial
interests or personal relationships that could have appeared to influence
the work reported in this paper.

\section*{Data Availability}

The \FixXPT{} source is distributed as a LAMMPS \code{XPT} package
(\url{https://github.com/atlas-nano/codes/tree/main/fix-xpt}) and archived on
Zenodo ~\cite{Ramji2026fixxpt}. The \pyxPT{} reference post-processor used for
cross-validation is openly available ~\cite{Ramji2026pyxpt}
(\url{https://github.com/atlas-nano/codes/tree/main/py-xPT}; version 1.0.1 archived at
\url{https://doi.org/10.5281/zenodo.22697489}). The LAMMPS input decks for the
liquid argon, SPC/E water and benzene benchmarks of Section~\ref{verification}, with the
reference output of the argon run, are included with the \FixXPT{} distribution.

\appendix
%% Appendix equations restart as (A.1); cas-sc does not do this itself.
\setcounter{equation}{0}
\renewcommand{\theequation}{A.\arabic{equation}}

\section{Desjarlais Memory-Function Model: Implementation}\label{app:desjarlais}

\subsection{Physical Motivation}\label{physical-motivation}

The standard 2PT Lorentzian gas DoS is derived by assuming that the
memory function of the generalized Langevin equation (GLE) for the
normalized VACF is a delta function in time, the Markovian limit. This
assumption works well for simple monatomic liquids and soft-sphere
fluids, but breaks down for systems with structured cages (liquid
metals, associated liquids, dense ionic systems), where the VACF
exhibits a backscattering shoulder or a slow secondary decay not
captured by a single Lorentzian.

Desjarlais ~\cite{Desjarlais2013} proposed replacing the Markovian memory kernel with a
Gaussian:

\begin{equation}
K_g(\tau) = A_g \exp\left(-B_g \tau^{2}\right)
\end{equation}

where $A_g$ sets the amplitude and $B_g$ \textgreater{} 0 controls the width of
the memory function. The key parameter is $B_g$: $B_g$ $\rightarrow$$\infty$ collapses
the Gaussian to a delta function, recovering the Lorentzian limit, while
finite $B_g$ introduces memory effects that broaden the gas DoS and
suppress the zero-frequency excess. The amplitude $A_g$ is not an
independent parameter; it is fixed by the zero-frequency constraint
$f_g F_g(0) = S_0/12N$, which gives $A_g = \left(f_g/S_0\right)\sqrt{4B_g/\pi}$ in the
code-scaled variables of Section~\ref{derivation-of-the-gas-dos}. This yields
a gas DoS that retains the correct normalization ($g_{gas}$(0) = $S_0$)
and integrates to $fN_{dof}$ degrees of freedom, while providing a more
flexible shape for matching the actual computed DoS.

\subsection{Derivation of the Gas DoS}\label{derivation-of-the-gas-dos}

The GLE with a Gaussian memory kernel has the Laplace-transform
solution:

\begin{equation}\label{eq:gle-solution}
F_g(\nu) = \frac{1}{2}\left[\frac{1}{K_g(i\omega)+i\omega} + \frac{1}{K_g(-i\omega)-i\omega}\right], \quad \omega = 2\pi \nu
\end{equation}

where $K_g(s)$is the Laplace transform of $K_g$($\tau$). For
the Gaussian kernel, this transform is given by:

\begin{equation}
K_g(i\omega)=A_g\sqrt{\dfrac{\pi}{4B_g}}\exp(-z^2)\,\mathrm{Erfc}(iz), \qquad z=\dfrac{\omega}{2\sqrt{B_g}}
\end{equation}

Using the Dawson-integral identity $\mathrm{Erfc}(iz)=1-\dfrac{2i}{\sqrt{\pi}}\exp(z^2)D(z)$, the
denominator of Eq.~\ref{eq:gle-solution} decomposes into real and imaginary parts:

\begin{equation}
\mathrm{Re}=A_g\sqrt{\dfrac{\pi}{4B_g}}\exp(-z^2), \qquad \mathrm{Im}=\omega-\dfrac{A_g}{\sqrt{B_g}}D(z)
\end{equation}

so that $F_g(\nu)=\mathrm{Re}/\left(\mathrm{Re}^2+\mathrm{Im}^2\right)$.

\textbf{Code-scaled variables.} The \FixXPT{} implementation works with the
code-normalized frequency $\tilde{\nu} = \pi\nu/6N$ (where $N$ is the
number of atoms or molecules) and the code parameter $B_g$ measured in
units of $\tilde{\nu}_0^{2}$, performing a change of variables
compared to the published work for computational convenience:
$\omega^{\mathrm{Desjarlais}} = 12N\tilde{\nu}$ and $B_g^{\mathrm{Desjarlais}}=(12N)^2 B_g$. With the further
substitution $y=\tilde{\nu}/\sqrt{4B_g}$, $A_g=(f/S_0)\sqrt{4B_g/\pi}$ and the
Dawson integral $D(y)=\exp(-y^2)\int_0^y \exp(t^2)\,dt$, the gas DoS in code units is:

\begin{equation}
g_{gas}^{HS}(\nu;B_g)=\dfrac{A_g\sqrt{\pi/(4B_g)}\exp(-y^2)f}{\left\{A_g^2\pi\exp(-2y^2)/4B_g\right\}+\left\{A_g^2D^2(y)/B_g-4A_g yD(y)+\tilde{\nu}^{2}\right\}}
\end{equation}

where the first $\{\ldots\}$ in the denominator is $\mathrm{Re}^2$ and the second $\{\ldots\}$ is
$\mathrm{Im}^2$.

\textbf{Boundary conditions.} At $\nu = 0$, $D(0) = 0$ and $e^{-y^2} = 1$,
so the denominator reduces to $A_g^2\pi/(4B_g)=(f/S_0)^2$ while the
numerator reduces to $f^2/S_0$, giving $g_{gas}^{HS}(0)=S_0$ by
construction. In
\FixXPT{}, this limit is handled explicitly to avoid a division by zero:

\code{if (nu == 0.0) return s0; // exact limit avoids 0/0}

\textbf{Lorentzian limit.} As $B_g \rightarrow \infty$ with $\nu$ fixed, $y \rightarrow 0$ and $D(y) \sim y$,
the $\mathrm{Im}^2$ term reduces to $\tilde{\nu}^{2}$, and the numerator
approaches $f^{2}/S_0$. The result is:

\begin{equation}
\lim_{B_g\to\infty} g_{gas}^{HS}(\nu)=\dfrac{S_0}{1+\left(\dfrac{\pi S_0\nu}{6Nf}\right)^2}
\end{equation}

which is exactly the standard Lorentzian (hard-sphere) gas DoS.

\subsection{Numerical Evaluation of the Dawson Integral}\label{numerical-evaluation-of-the-dawson-integral}

The Dawson integral $D(y)=\exp(-y^2)\int_0^y \exp(t^2)\,dt$ must be evaluated
at $y=\tilde{\nu}/\sqrt{4B_g}=\pi\nu/\left(12N\sqrt{B_g}\right)$ for every frequency bin. \FixXPT{} uses two
regimes:

\begin{itemize}
\item
  \textbf{Small argument} ($|y| < 0.2$): the truncated Taylor series
  \(D(y) = y\,[\,1 - \tfrac{2}{3}y^{2}(1 - \tfrac{2}{5}y^{2}(1 - \tfrac{2}{7}y^{2}))\,]\), which avoids the \(0/0\) cancellation of the
  defining integral near the origin.
\item
  \textbf{All other arguments:} Rybicki's method~\cite{Rybicki1989} (Numerical Recipes~\cite{Press1992} \S{}6.10),
  which evaluates \(D(y)\) from a short sum of exponentially spaced
  sample points with step \(H = 0.4\) and six tabulated weights
  \(c_{i} = \exp[-((2i-1)H)^{2}]\), \(i = 1\ldots6\). This is accurate to
  \(\sim 10^{-8}\) across the entire real axis and matches
  \code{scipy.special.dawsn} (the routine used by the \pyxPT{} reference
  implementation).
\end{itemize}

Rybicki's method is uniformly accurate, adds no external-library
dependency, and requires only six exponential evaluations per call (in
the memory-function path only). A piecewise rational-exponential
approximation of the same integral carries a mid-range error of
$\sim$0.6\%, too large for the agreement with the reference
post-processor reported in Section~\ref{verification}.

\subsection{\texorpdfstring{Self-Consistent Determination of $B_g$}{Self-Consistent Determination of Bg}}\label{self-consistent-determination-of-bg}

The shape parameter $B_g$ is determined by requiring that
$g_{gas}^{HS}$ matches the actual computed DoS at the ``tail'' of the spectrum, the last frequency bin before the DoS drops into the
noise floor:

\begin{lstlisting}[language=C++]
double noise = 1e-6 * s0;
int j = (int)pwr.size() - 1;
while (j > 0 && pwr[j] < noise) j--;   // walk back from Nyquist
double nu_tail = j * dnu;
double target  = pwr[j];               // actual DoS at tail
\end{lstlisting}

The bisection finds $B_g$ $\in$ {[}$B_{lo}$, $B_{hi}${]} such that
$g_{gas}^{HS}\left(\nu_{max},B_g\right)$ = target to relative tolerance $10^{-8}$ in at most
80 iterations:

\begin{lstlisting}[language=C++]
for (int it = 0; it < 80; it++) {
  double Bg_mid = 0.5*(Bg_lo + Bg_hi);
  (sgmf_des(nu_tail, s0, f_g, Bg_mid, nmol) < target
     ? Bg_lo : Bg_hi) = Bg_mid;
  if ((Bg_hi - Bg_lo)/(Bg_hi + Bg_lo) < 1e-8) break;
}
return 0.5*(Bg_lo + Bg_hi);
\end{lstlisting}

The initial bracket is {[}$10^{-3}$, 1{]}, expanded geometrically if the
target lies outside: $B_{lo}$ is halved until $g_{gas}^{HS}\left(B_{lo}\right)$$\leq$
target, and $B_{hi}$ is doubled until $g_{gas}^{HS}\left(B_{hi}\right)$$\geq$ target. If
$j = 0$ (i.e., all bins are below the noise floor) or f $\leq$ 0, the
function returns $B_g$ = 0 and the standard Lorentzian is used as a
fallback.

\textbf{Bracketing.} The bisection assumes a single increasing branch,
which holds over most of the useful range: larger $B_g$ means a narrower
Gaussian memory kernel $K_g(\tau)$ in time, i.e.\ shorter-lived memory,
and in the limit $B_g \rightarrow \infty$ the kernel collapses to a delta
function and the gas DoS approaches the Lorentzian. The approach to that
limit is not monotonic, however. At fixed $\nu_{tail}$,
$g_{gas}^{HS}(\nu;B_g)$ rises from zero, overshoots the Lorentzian value
at finite $B_g$, and relaxes back to it from above; note that
$A_g/\sqrt{B_g} = 2f/(S_0\sqrt{\pi})$ is independent of $B_g$, so the
$B_g$ dependence resides entirely in the Dawson term. The overshoot is
slight in the regime the tail fit occupies (0.3--5\% for the liquid
argon and water windows reported here), but it has two consequences. A
target lying between the Lorentzian limit and the local maximum admits
two roots, and the bisection returns whichever the initial bracket
encloses; a target above the maximum is unreachable, and the doubling
of $B_g^{hi}$ terminates at its $10^{8}$ cap rather than at a fit. The
$\min(g_{gas}, g)$ ceiling of the following section bounds the effect on
the partition in both cases.

\subsection{Integration with the Analysis Pipeline}\label{integration-with-the-analysis-pipeline}

Once $B_g$ is determined, the gas DoS array is constructed by the
dispatcher \code{gas\_comp}, which switches between Lorentzian ($B_g$ = 0)
and Memory-function ($B_g$ \textgreater{} 0) automatically:

\begin{lstlisting}[language=C++]
double FixXPT::gas_comp(int j, double dnu, double dos_j,
                        double s0_c, double f_c,
                        double Bg_c, double nmol_c) const
{
  double nu = j * dnu;
  double sg = (Bg_c > 0.0)
    ? ((j == 0) ? s0_c : sgmf_des(nu, s0_c, f_c, Bg_c, nmol_c))
    : ((nu == 0.0) ? s0_c
                   : s0_c / (1.0 + pow(PI*s0_c*nu/(6.0*nmol_c*f_c), 2.0)));
  return (sg > dos_j) ? dos_j : sg;   // cap at actual DoS
}
\end{lstlisting}

The ceiling \code{min(sg, dos\_j)} prevents the gas DoS from exceeding the
actual DoS at any bin, a physical requirement since $g_{gas}$ $\leq$ $g$
everywhere. The solid DoS is then $g_{solid}(j) = g(j) - g_{gas}(j)$ $\geq$
0 by construction. All subsequent thermodynamic integration proceeds
identically for the \kw{rigorous} and \kw{desjarlais} refinements; the distinction is entirely in stored
value of the DoS\textsubscript{gas}($\omega$) subsystem.

In molecular mode, \code{refine\_Bg\_des} is called separately for the
translational and rotational DoS components, yielding independent $B_g$
values for each. The vibrational component is always pure-solid and
bypasses the 2PT procedure.

\bibliographystyle{elsarticle-num-names}
\bibliography{refs}

@article{Lin2003,
  author = {Lin, Shiang-Tai and Blanco, Mario and Goddard III, William A.},
  title = {The two-phase model for calculating thermodynamic properties of liquids from molecular dynamics: Validation for the phase diagram of {L}ennard-{J}ones fluids},
  journal = {J. Chem. Phys.},
  volume = {119},
  pages = {11792--11805},
  year = {2003},
  doi = {10.1063/1.1624057}
}

@article{pascal2012hydrophobic,
  title={Hydrophobic segregation, phase transitions and the anomalous thermodynamics of water/methanol mixtures},
  author={Pascal, Tod A and Goddard III, William A},
  journal={The Journal of Physical Chemistry B},
  volume={116},
  number={47},
  pages={13905--13912},
  year={2012},
  publisher={ACS Publications},
  doi={10.1021/jp309693d}
}

@article{Lin2010,
  author = {Lin, Shiang-Tai and Maiti, Prabal K. and Goddard III, William A.},
  title = {Two-phase thermodynamic model for efficient and accurate absolute entropy of water from molecular dynamics simulations},
  journal = {J. Phys. Chem. B},
  volume = {114},
  pages = {8191--8198},
  year = {2010},
  doi = {10.1021/jp103120q}
}

@article{Desjarlais2013,
  author = {Desjarlais, Michael P.},
  title = {First-principles calculation of entropy for liquid metals},
  journal = {Phys. Rev. E},
  volume = {88},
  pages = {062145},
  year = {2013},
  doi = {10.1103/PhysRevE.88.062145}
}

@article{Zwanzig1954,
  author = {Zwanzig, Robert W.},
  title = {High-temperature equation of state by a perturbation method. {I}. {N}onpolar gases},
  journal = {J. Chem. Phys.},
  volume = {22},
  pages = {1420--1426},
  year = {1954},
  doi = {10.1063/1.1740409}
}

@article{Kirkwood1935,
  author = {Kirkwood, John G.},
  title = {Statistical mechanics of fluid mixtures},
  journal = {J. Chem. Phys.},
  volume = {3},
  pages = {300--313},
  year = {1935},
  doi = {10.1063/1.1749657}
}

@article{Carnahan1969,
  author = {Carnahan, Norman F. and Starling, Kenneth E.},
  title = {Equation of state for nonattracting rigid spheres},
  journal = {J. Chem. Phys.},
  volume = {51},
  pages = {635--636},
  year = {1969},
  doi = {10.1063/1.1672048}
}

@article{Endo2022,
  author = {Endo, Takatsugu and Sunada, Kouki and Sumida, Hiroki and Kimura, Yoshifumi},
  title = {Origin of low melting point of ionic liquids: dominant role of entropy},
  journal = {Chem. Sci.},
  volume = {13},
  pages = {7560--7565},
  year = {2022},
  doi = {10.1039/D2SC02342C}
}

@article{Pascal2011PNAS,
  author = {Pascal, Tod A. and Goddard III, William A. and Jung, Yousung},
  title = {Entropy and the driving force for the filling of carbon nanotubes with water},
  journal = {Proc. Natl. Acad. Sci. U.S.A.},
  volume = {108},
  pages = {11794--11798},
  year = {2011},
  doi = {10.1073/pnas.1108073108}
}

@article{Thompson2022,
  author = {Thompson, Aidan P. and Aktulga, H. Metin and Berger, Richard and Bolintineanu, Dan S. and Brown, W. Michael and Crozier, Paul S. and in~'t~Veld, Pieter J. and Kohlmeyer, Axel and Moore, Stan G. and Nguyen, Trung Dac and Shan, Ray and Stevens, Mark J. and Tranchida, Julien and Trott, Christian and Plimpton, Steven J.},
  title = {{LAMMPS} -- a flexible simulation tool for particle-based materials modeling at the atomic, meso, and continuum scales},
  journal = {Comput. Phys. Commun.},
  volume = {271},
  pages = {108171},
  year = {2022},
  doi = {10.1016/j.cpc.2021.108171}
}

@article{Sun2017,
  author = {Sun, Tao and Xian, Jiawei and Zhang, Huai and Zhang, Zhigang and Zhang, Yigang},
  title = {Two-phase thermodynamic model for computing entropies of liquids reanalyzed},
  journal = {J. Chem. Phys.},
  volume = {147},
  pages = {194505},
  year = {2017},
  doi = {10.1063/1.5001798}
}

@article{Buarque2026,
  author = {Buarque, Ricardo and Gascon, Mauro and Pascal, Tod A.},
  title = {An anharmonic liquid-entropy functional from the {M}ori--{Z}wanzig memory kernel},
  journal = {arXiv:2607.29026 [physics.chem-ph]},
  year = {2026},
  doi = {10.48550/arXiv.2607.29026}
}

@misc{Ramji2026pyxpt,
  author = {Ramji, Robert S. and Pascal, Tod A.},
  title = {{py-xPT}: Two- and three-phase thermodynamics ({2PT}/{3PT}) entropy and free energy from molecular dynamics},
  version = {1.0.1},
  note = {Version 1.0.1},
  publisher = {Zenodo},
  year = {2026},
  doi = {10.5281/zenodo.22697489}
}

@misc{Ramji2026fixxpt,
  author = {Ramji, Robert S. and Buarque, Ricardo and Ruan, Xiaoxu and Pascal, Tod A.},
  title = {{FixXPT}: On-the-fly two- and three-phase thermodynamics for {LAMMPS}},
  version = {1.0.1},
  note = {Version 1.0.1},
  publisher = {Zenodo},
  year = {2026},
  doi = {10.5281/zenodo.22729714}
}

@misc{Ramji20232pt,
  author       = {Ramji, Robert and Pascal, Tod},
  title        = {2PT post trajectory analysis code},
  version      = {v1.6},
  year         = {2023},
  month        = mar,
  publisher    = {Zenodo},
  doi          = {10.5281/zenodo.7731073},
  url          = {https://doi.org/10.5281/zenodo.7731073}
}

@article{ramirez2010efficient,
  title={Efficient on the fly calculation of time correlation functions in computer simulations},
  author={Ram{\'\i}rez, Jorge and Sukumaran, Sathish K and Vorselaars, Bart and Likhtman, Alexei E},
  journal={The Journal of chemical physics},
  volume={133},
  number={15},
  year={2010},
  publisher={AIP Publishing}
}

@article{Wang2018,
  author  = {Wang, Han and Zhang, Linfeng and Han, Jiequn and E, Weinan},
  title   = {{DeePMD-kit}: A deep learning package for many-body potential energy representation and molecular dynamics},
  journal = {Comput. Phys. Commun.},
  volume  = {228},
  pages   = {178--184},
  year    = {2018},
  doi     = {10.1016/j.cpc.2018.03.016}
}

@article{Zhang2018,
  author  = {Zhang, Linfeng and Han, Jiequn and Wang, Han and Car, Roberto and E, Weinan},
  title   = {Deep Potential Molecular Dynamics: A Scalable Model with the Accuracy of Quantum Mechanics},
  journal = {Phys. Rev. Lett.},
  volume  = {120},
  pages   = {143001},
  year    = {2018},
  doi     = {10.1103/PhysRevLett.120.143001}
}

@article{Berendsen1987,
  author  = {Berendsen, H. J. C. and Grigera, J. R. and Straatsma, T. P.},
  title   = {The missing term in effective pair potentials},
  journal = {J. Phys. Chem.},
  volume  = {91},
  pages   = {6269--6271},
  year    = {1987},
  doi     = {10.1021/j100308a038}
}

@article{Abascal2005,
  author  = {Abascal, J. L. F. and Vega, C.},
  title   = {A general purpose model for the condensed phases of water: {TIP4P/2005}},
  journal = {J. Chem. Phys.},
  volume  = {123},
  pages   = {234505},
  year    = {2005},
  doi     = {10.1063/1.2121687}
}

@article{Wang2004,
  author  = {Wang, Junmei and Wolf, Romain M. and Caldwell, James W. and Kollman, Peter A. and Case, David A.},
  title   = {Development and testing of a general amber force field},
  journal = {J. Comput. Chem.},
  volume  = {25},
  pages   = {1157--1174},
  year    = {2004},
  doi     = {10.1002/jcc.20035}
}

@article{Jorgensen1996,
  author  = {Jorgensen, William L. and Maxwell, David S. and Tirado-Rives, Julian},
  title   = {Development and Testing of the {OPLS} All-Atom Force Field on Conformational Energetics and Properties of Organic Liquids},
  journal = {J. Am. Chem. Soc.},
  volume  = {118},
  pages   = {11225--11236},
  year    = {1996},
  doi     = {10.1021/ja9621760}
}

@article{Sutton1990,
  author  = {Sutton, A. P. and Chen, J.},
  title   = {Long-range {Finnis--Sinclair} potentials},
  journal = {Philos. Mag. Lett.},
  volume  = {61},
  pages   = {139--146},
  year    = {1990},
  doi     = {10.1080/09500839008206493}
}

@article{Rybicki1989,
  author  = {Rybicki, George B.},
  title   = {{Dawson's} Integral and the Sampling Theorem},
  journal = {Comput. Phys.},
  volume  = {3},
  number  = {2},
  pages   = {85--87},
  year    = {1989},
  doi     = {10.1063/1.4822832}
}

@book{Press1992,
  author    = {Press, William H. and Teukolsky, Saul A. and Vetterling, William T. and Flannery, Brian P.},
  title     = {Numerical Recipes in {C}: The Art of Scientific Computing},
  edition   = {2nd},
  publisher = {Cambridge University Press},
  address   = {Cambridge},
  year      = {1992}
}

@book{Chapman1970,
  author    = {Chapman, Sydney and Cowling, T. G.},
  title     = {The Mathematical Theory of Non-uniform Gases},
  edition   = {3rd},
  publisher = {Cambridge University Press},
  address   = {Cambridge},
  year      = {1970}
}

@article{Falcone1967,
  author  = {Falcone, D. R. and Douglass, Dean C. and McCall, David W.},
  title   = {Self-diffusion in benzene},
  journal = {J. Phys. Chem.},
  volume  = {71},
  number  = {8},
  pages   = {2754--2755},
  year    = {1967},
  doi     = {10.1021/j100867a067}
}

@article{Trott2022,
  author  = {Trott, Christian R. and Lebrun-Grandi{\'e}, Damien and Arndt, Daniel and Ciesko, Jan and Dang, Vinh and Ellingwood, Nathan and Gayatri, Rahulkumar and Harvey, Evan and Hollman, Daisy S. and Ibanez, Dan and Liber, Nevin and Madsen, Jonathan and Miles, Jeff and Poliakoff, David and Powell, Amy and Rajamanickam, Sivasankaran and Simberg, Mikael and Sunderland, Dan and Turcksin, Bruno and Wilke, Jeremiah},
  title   = {{Kokkos} 3: Programming Model Extensions for the Exascale Era},
  journal = {IEEE Trans. Parallel Distrib. Syst.},
  volume  = {33},
  pages   = {805--817},
  year    = {2022},
  doi     = {10.1109/TPDS.2021.3097283}
}

@article{Schaetzel1987,
  author  = {Sch{\"a}tzel, Klaus},
  title   = {Correlation techniques in dynamic light scattering},
  journal = {Appl. Phys. B},
  volume  = {42},
  pages   = {193--213},
  year    = {1987},
  doi     = {10.1007/BF00693937}
}

@article{Magatti2001,
  author  = {Magatti, Davide and Ferri, Fabio},
  title   = {Fast multi-tau real-time software correlator for dynamic light scattering},
  journal = {Appl. Opt.},
  volume  = {40},
  pages   = {4011--4021},
  year    = {2001},
  doi     = {10.1364/AO.40.004011}
}

@article{Wiener1930,
  author  = {Wiener, Norbert},
  title   = {Generalized harmonic analysis},
  journal = {Acta Math.},
  volume  = {55},
  pages   = {117--258},
  year    = {1930},
  doi     = {10.1007/BF02546511}
}

@article{Khintchine1934,
  author  = {Khintchine, A.},
  title   = {Korrelationstheorie der station{\"a}ren stochastischen Prozesse},
  journal = {Math. Ann.},
  volume  = {109},
  pages   = {604--615},
  year    = {1934},
  doi     = {10.1007/BF01449156}
}

@book{Hockney1988,
  author    = {Hockney, R. W. and Eastwood, J. W.},
  title     = {Computer Simulation Using Particles},
  publisher = {Adam Hilger},
  address   = {Bristol},
  year      = {1988},
  doi       = {10.1201/9780367806934}
}

@article{Hu2023,
  author  = {Hu, Taiping and Dai, Fu-zhi and Zhou, Guobing and Wang, Xiaoxu and Xu, Shenzhen},
  title   = {Unraveling the Dynamic Correlations between Transition Metal Migration and the Oxygen Dimer Formation in the Highly Delithiated {Li$_x$CoO$_2$} Cathode},
  journal = {J. Phys. Chem. Lett.},
  volume  = {14},
  pages   = {3677--3684},
  year    = {2023},
  doi     = {10.1021/acs.jpclett.3c00506}
}

@article{Zeron2019,
  author  = {Zeron, I. M. and Abascal, J. L. F. and Vega, C.},
  title   = {A force field of {Li$^+$}, {Na$^+$}, {K$^+$}, {Mg$^{2+}$}, {Ca$^{2+}$}, {Cl$^-$}, and {SO$_4^{2-}$} in aqueous solution based on the {TIP4P/2005} water model and scaled charges for the ions},
  journal = {J. Chem. Phys.},
  volume  = {151},
  pages   = {134504},
  year    = {2019},
  doi     = {10.1063/1.5121392}
}

@article{deLucas2024,
  author  = {de Lucas, Marcos and Blazquez, Samuel and Troncoso, Jacobo and Vega, Carlos and G{\'a}mez, Francisco},
  title   = {Dressing a Nonpolarizable Force Field for {OH$^-$} in {TIP4P/2005} Aqueous Solutions with Charge},
  journal = {J. Phys. Chem. Lett.},
  volume  = {15},
  pages   = {9411--9418},
  year    = {2024},
  doi     = {10.1021/acs.jpclett.4c02261}
}

@article{Yeh2004,
  author  = {Yeh, In-Chul and Hummer, Gerhard},
  title   = {System-Size Dependence of Diffusion Coefficients and Viscosities from Molecular Dynamics Simulations with Periodic Boundary Conditions},
  journal = {J. Phys. Chem. B},
  volume  = {108},
  pages   = {15873--15879},
  year    = {2004},
  doi     = {10.1021/jp0477147}
}

\end{document}